%% file: inner_bonds_v2.tex
\documentclass[twocolumn,aps,prx,longbibliography,superscriptaddress]{revtex4-2}
\usepackage{amsmath,amssymb}
\usepackage{graphicx}
\usepackage{hyperref}
\usepackage[normalem]{ulem}
\hypersetup{colorlinks=true, linkcolor=blue, citecolor=blue, urlcolor=blue}
\usepackage{bbold}

\begin{document}
\author{Moritz Cygorek}
\affiliation{SUPA, Institute of Photonics and Quantum Sciences, Heriot-Watt University, Edinburgh, EH14 4AS, United Kingdom}
\affiliation{Condensed Matter Theory, Technical University of Dortmund, 44227 Dortmund, Germany}
\author{Erik M. Gauger}
\affiliation{SUPA, Institute of Photonics and Quantum Sciences, Heriot-Watt University, Edinburgh, EH14 4AS, United Kingdom}
\title{Understanding and utilizing the inner bonds of process tensors}

\begin{abstract}
Process tensor matrix product operators (PT-MPOs) enable numerically exact simulations for an unprecedentedly broad range of open quantum systems.
By representing environment influences in MPO form, they can be efficiently compressed using established algorithms. 
The dimensions of inner bonds of the compressed PT-MPO may be viewed as an indicator of the complexity of the environment. 
Here, we show that the inner bonds themselves, not only their dimensions, have a concrete physical meaning:
They represent the subspace of the full environment Liouville space
which hosts environment excitations that may influence the subsequent open quantum
systems dynamics the most.
This connection can be expressed in terms of lossy linear transformations,
whose pseudoinverses facilitate the extraction of environment observables.
We demonstrate this by extracting the environment spin of a central spin problem,
the current through a quantum system coupled to two leads, the number of photons 
emitted from quantum emitters into a structured environment, and the distribution
of the total absorbed energy in a driven non-Markovian quantum system into system, environment,
and interaction energy terms.
Numerical tests further indicate that different PT-MPO algorithms compress environments to similar subspaces.
Thus, the physical interpretation of inner bonds of PT-MPOs both provides
a conceptional understanding and it enables new practical applications. 
\end{abstract}

\maketitle
\section{Introduction}
All real-world quantum systems are invariably coupled to their surrounding environment. 
When the coupling is weak, the Born-Markov approximation
provides a sufficient description of the open quantum systems dynamics
in the form of Lindblad master equations~\cite{BreuerPetruccione}.
However, in many cases, as in charge or excitation transfer in 
chemical~\cite{Scholes_natchem2021} or biological 
systems~\cite{Chin_noise-assisted}, 
in spin systems~\cite{DMS_Thurn,DMS_exciton}, 
in solid-state quantum emitters~\cite{variationalPRB,
Kaldewey2017, Reiter_review2019, Denning_decoupling},
or superconducting qubits~\cite{PT_White2020,Gulacsi2023},
the system-environment coupling is strong enough to lead to sizable
non-Markovian memory effects. 
Because the Born-Markov approximation becomes insufficient in such situations, 
predicting the dynamics of open quantum systems then requires methods that 
accurately account for effects such as renormalization of system energy scales~\cite{Review_Nazir,
PI_entangled_PRL}, 
environment-assisted transitions~\cite{Chin_noise-assisted,PI_singlephoton,
PI_entangled_PRB},
non-exponential decay~\cite{RevModPhys_Leggett,CoopWiercinski}, 
the formation of quasi-particles like polarons~\cite{Review_Nazir},
and deviations of multi-time correlation functions from predictions
based on the quantum regression theorem~\cite{PI_QRT}.

\begin{figure*}
\includegraphics[width=0.99\textwidth]{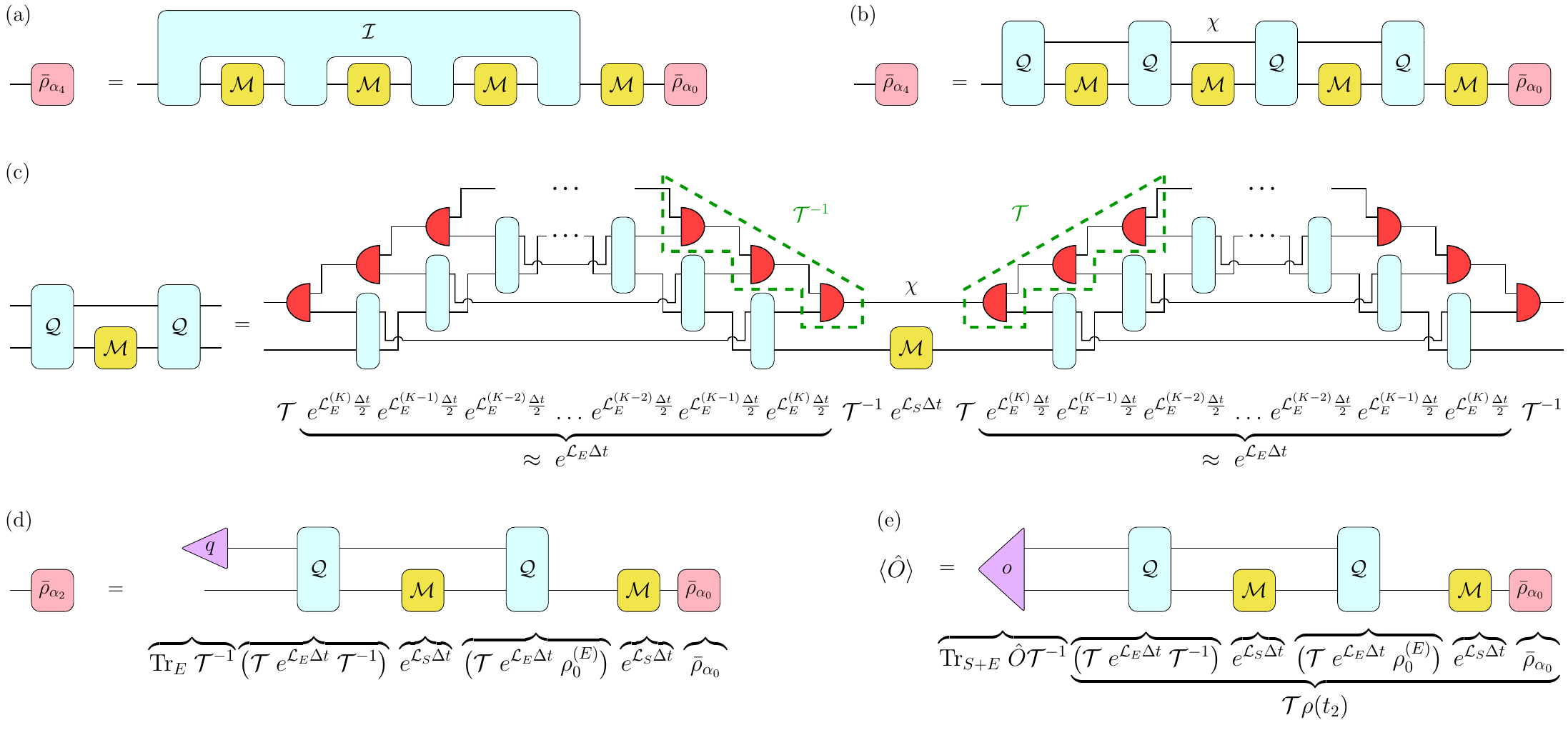}
\caption{\label{fig:sketch}%
(a) Time-discretized Feynman-Vernon path integral for the 
reduced system density matrix after $n=4$ time steps $\bar{\rho}_{\alpha_4}$, 
which involves the free system propagator
$\mathcal{M}=e^{\mathcal{L}_S \Delta t}$ and the influence functional 
$\mathcal{I}$. 
The comb-like geometry is required to account for the time-non-local information flow
through non-Markovian (finite-memory) environments.
(b) The PT-MPO representation of the influence functional with matrices $\mathcal{Q}$
facilitates easy evaluation of the same path integral as a sequence of matrix multiplications. 
The non-local information flow is now mediated by the inner bonds with maximal dimension~$\chi$.
(c)~Section of a PT-MPO obtained using \emph{Automated Compression of 
Environments} (ACE)~\cite{ACE}: A PT-MPO for an environment composed of 
$K$ non-interacting modes is constructed by combining and compressing 
PT-MPOs of the individual modes one by one. The PT-MPO of a single mode $k$ 
is generated by the $k$-th environment Liouvillian 
$e^{\mathcal{L}_E^{(k)} \Delta t}$, 
where the inner bonds span the $k$-th environment Liouville space.
A central finding of this article is that the matrices $\mathcal{Q}$ of the 
overall PT-MPO can be explicitly linked to the full environment Liouville 
propagator via
$\mathcal{Q}=\mathcal{T}e^{\mathcal{L}_E \Delta t}\mathcal{T}^{-1}$, 
where $\mathcal{T}$ are lossy (rank-reducing) transformations and 
$\mathcal{T}^{-1}$ are their pseudoinverses. These are themselves of 
MPO form.
With this insight, the utility of the inner bonds is not limited to 
truncating PT-MPOs at earlier time steps (d); It also enables the
extraction of environment and mixed system-environment observables (e).
}
\end{figure*}

Process tensor matrix product operators (PT-MPOs) provide an attractive and practical solution to tackle such challenging problems. They can be understood as efficient representations of Feynman-Vernon influence 
functionals~\cite{FeynmanVernon} [see Fig.~\ref{fig:sketch}(a)]
in the form of matrix product operators (MPOs)~\cite{MPS_Orus,MPS_Schollwoeck} 
[see Fig.~\ref{fig:sketch}(b)] or tensor trains~\cite{TensorTrain}.
PT-MPOs can be used to simulate open quantum systems numerically 
exactly, i.e.~they include all effects generated by the 
microscopic system and environment Hamiltonians to all orders in the coupling.
Any inaccuracy is then purely the result of insufficient numerical convergence and can, in principle, be made arbitrarily small by choosing more stringent convergence parameters.

Owing to their practicality and efficiency, PT-MPOs have seen wide adoption 
within a few years of their inception~\cite{PT_PRA,JP}. For example,
Denning \textit{et al.}~\cite{Denning_decoupling},  
Fux \textit{et al.}~\cite{Fux_OptimalControl}, and Vannucci and Gregersen~\cite{Vannucci2023} have 
used PT-MPOs to study the dynamics of semiconductor quantum dots interacting with a
phonon bath, while Richter and Hughes have described two
emitters coupled to a common waveguide~\cite{PRL_Richter}. 
We recently demonstrated a divide-and-conquer algorithm for constructing periodic PT-MPOs~\cite{DnC}, which
has enabled million-time-step simulations, e.g., for modelling experiments measuring
quantum dot emission spectra after time-dependent (pulsed) driving by
Boos \textit{et al.}~\cite{Boos_spectra} as well as the analysis of two-color excitation with strongly
off-resonant laser pulses by Bracht \textit{et al.}~\cite{Bracht_SUPER_biexc}.
Link, Tu, and Strunz described a method to calculate periodic PT-MPOs
with linear scaling with respect to memory time~\cite{Link_infinite}.
Impurity problems have been adressed by 
Abanin \textit{et al.}~\cite{PT_Abanin} and 
Reichman \textit{et al.}~\cite{PT_Reichman} using fermionic PT-MPOs.
Examples involving PT-MPOs for interacting boson and spin environments
have been given by Ye and Chan~\cite{PT_Ye} 
and Guo \textit{et al.}~\cite{PT_Guo}, respectively.
Moreover, the PT-MPO formalism remains numerically exact when open quantum systems 
coupled to multiple environments are simulated by interleaving the corresponding PT-MPOs, which are constructed independently of each other.
This has been exploited for investigations of non-additive effects of 
two non-Markovian baths~\cite{twobath} as well as of cooperative effects in
multi-quantum-emitter systems~\cite{CoopWiercinski}, 
and also paves the way for scalable 
numerically exact simulations of quantum networks~\cite{Fux_spinchain}.
With the algorithm
\textit{Automated Compression of Environments} (ACE)~\cite{ACE},
we have recently extended the scope of process tensor methods to environments 
with arbitrary Hamiltonians composed of independent modes.
Despite the emerging broad utility and practical power of PT-MPOs, an understanding of their inner structure and information content remains lacking. 

Originally, process tensors (PTs) were introduced 
by Pollock \emph{et al.}~in Ref.~\cite{PT_PRA} as a means to characterize non-Markovian 
quantum processes. As such, they generalize quantum channels, 
which describe how quantum systems change over a fixed time interval,
to a description spanning multiple time steps. To tackle the exponential growth of PTs 
with the number of time steps, the PT---a one-dimensional object in time---is
mapped to the state of a spatially one-dimensional many-body quantum system 
via a generalized Choi-Jamio{\l}kowski state-channel 
isomorphism~\cite{Choi, Jamiolkowski}. This Choi state is then efficiently 
represented as a matrix product operator (MPO). 
Contemporarily, Strathearn \emph{et al.}~\cite{TEMPO} developed TEMPO,
which uses MPOs to improve the performance and resolvable memory duration of the QUAPI~\cite{QUAPI1,QUAPI2} 
variant of Feynman-Vernon~\cite{FeynmanVernon} path integral simulations. 
The fact that PTs are equivalent to Feynman-Vernon influence functionals 
for spin-boson-type environments was observed by J{\o}rgensen and Pollock in Ref.~\cite{JP}.
There, it was also found that TEMPO corresponds to a particular contraction of 
a tensor network that also describes the PT-MPO, and that it is typically
more efficient to directly calculate the latter.
Thereby, Ref.~\cite{JP} has laid the grounds for
using PT-MPOs as practical tools to simulate open quantum systems, 
beyond its original purpose for characterizing unknown non-Markovian
quantum processes~\cite{PT_PRA,PT_White2020}.

Invariably, the dimensions $\chi$ of the inner MPO bonds 
[see Fig.~\ref{fig:sketch}(b)] play a crucial role, as they are a key factor
for the required computational resources, e.g., for storing the PT-MPO in memory and performing operations with it.
Formally, the bond dimensionality can be linked to the R{\'e}nyi entropy of the Choi state \cite{Luchnikov}, borrowing from one-dimensional many-body MPO theory~\cite{VerstraeteCirac2006}.
Further, recent work provided an analytic bound for the PT-MPO bond dimension for an Ohmic spin-boson environment at zero temperature \cite{Abanin_bound}. There exists broad consensus~\cite{PT_PRA,Luchnikov,Abanin_bound,DnC} 
on the interpretation of the bond dimension $\chi$ as a measure for 
the simulation complexity of an open quantum system, with 
some authors also seeking links between $\chi$ and notions
of non-Markovianity~\cite{PT_Markov}, or of quantum chaos~\cite{PT_Chaos}.

By constrast, the inner bonds of many-body matrix product states are known to contain more physical information than just their dimensions. Most notably, the entanglement spectrum~\cite{entanglement_spectrum} as well as the transformation behavior of inner bonds of matrix product states can be used to identify topological order~\cite{Pollmann2010}.
Whether or not the inner bonds of PT-MPOs also contain meaningful physical information beyond their dimensions is not obvious from the derivation of PT-MPO algorithms based on Feynman-Vernon path integral expressions~\cite{JP, DnC,Link_infinite}. There, by construction, the inner bonds are used to encode long-range temporal correlations of the environment.
This is different in the ACE algorithm~\cite{ACE}, where the PT-MPO is constructed directly from the microscopic Hamiltonian
of a general environment composed of mutually noninteracting modes. PT-MPOs for individual modes are obtained by exponentiating their respective Liouvillians, before these individual PT-MPOs are combined and compressed to form the PT-MPO for the full environment [see Fig.~\ref{fig:sketch}(c)].
Before compression, the inner bonds of the individual mode PT-MPOs simply denote the Liouville space, i.e.~the squared Hilbert space, of the respective environment mode, which carries complete information about the state of the latter.
It can thus be expected that some of this information survives MPO compression.

In this article, we provide a precise interpretation of the meaning of the inner bonds of PT-MPOs and thereby of the PT-MPOs themselves.
By tracking how the inner bonds are transformed in every step of the 
ACE algorithm~\cite{ACE}, we identify the connection
between the inner bonds of the final (compressed) PT-MPO and the Liouville space of 
the environment in terms of lossy linear transformations
[see Fig.~\ref{fig:sketch}(c)].
Along with the explicit transformation matrices $\mathcal{T}$ 
(which vary from time step to time step), we obtain corresponding
pseudoinverses $\mathcal{T}^{-1}$.
Our main result is a conceptually clear interpretation of PT-MPOs:
The matrices $\mathcal{Q}$ forming the bulk of the PT-MPO 
\begin{align}
\label{eq:QTeT}
\mathcal{Q}=\mathcal{T} e^{\mathcal{L}_E \Delta t} \mathcal{T}^{-1},
\end{align}
are simply the propagators $e^{\mathcal{L}_E \Delta t}$ of the full environment
over a time step $\Delta t$ compressed to the relevant (contemporary) subspace via
$\mathcal{T}$ and $\mathcal{T}^{-1}$, where $\mathcal{L}_E$ is the 
environment Liouvillian in the superoperator notation described in 
section~\ref{sec:notation}.
This relevant subspace is implicitly and automatically determined by MPO compression.

As a consequence, the inner bonds carry time-local information about the dynamical
evolution of the environment, which can be partially reconstructed 
with the help of the pseudoinverses $\mathcal{T}^{-1}$
as indicated in Fig.~\ref{fig:sketch}(e). 
The extraction of environment and mixed system-environment observables
$\mathcal{O}$ is achieved by terminating the PT-MPO path sum with 
an object $o$, which we call an \emph{observable closure} and which expresses the action of 
$o \;\cdot =\textrm{Tr}_{S+E}\big\{\mathcal{O}\mathcal{T}^{-1} \cdot \big\}$, where 
$\textrm{Tr}_{S+E}$ denotes the trace over system and environment 
degrees of freedom.
In order to avoid the explicit storage of the matrices $\mathcal{T}^{-1}$ 
at each time step, we instead transform $o$ along the individual steps of
the PT-MPO construction in the ACE algorithm.

However, because the transformation  $\mathcal{T}$ is lossy, there is no guarantee that
a given environment observable can indeed be accurately reconstructed.
Comparing convergence of different environment observables therefore
allows one to probe what information is conveyed in the inner bonds 
of PT-MPOs and what is eliminated by MPO compression. 
The comparison with a hierarchy of Heisenberg equations of motion for
operator averages reveals that first order system-environment correlations
that directly enter the Heisenberg equation of motion for system observables are
much more accurately reproduced than environment observables further
down the hierarchy that 
affect the system only indirectly via influencing the first order correlations.
We shall presently demonstrate that
this insight can be utilized to devise alternative, more accurate schemes 
to extract environment observables from PT-MPO simulation.

We consider a variety of examples showcasing available access to environmental and mixed observables. 
First,  on the 
example of a central spin model with total spin conservation, we show that
the total environment spin can be faithfully obtained from the inner bonds 
of PT-MPOs. Next, we demonstrate the extraction of currents through a central
site coupled to two metallic leads at different chemical potentials. 
The convergence of environment observables is then analyzed in more detail 
on the example of photon emission from a quantum emitter. Finally, we 
calculate the total energy absorbed by an externally driven 
non-Markovian open quantum system as well as its distribution 
into terms associated to only the system, to only the environment, 
as well as to the system-environment interaction term, which is proportional
to system-environment correlations.
On the last example, we also perform a numerical experiment where we take the 
observable closures obtained from ACE simulations and, after fixing the
gauge freedom, apply them to inner bonds of PT-MPOs calculated using the
algorithm by J{\o}rgensen and Pollock~\cite{JP}, which is based on the 
Feynman-Vernon path integral expression for the influence functional
and never makes reference to any particular environment Liouville space. 
Yet, we find a remarkable agreement between the extracted environment
observables from both PT-MPOs, indicating that the space described by 
the inner bonds is essentially independent of the PT-MPO algorithm, and
hence a universal property of PT-MPOs.

This article is structured as follows: 
First, in Section~\ref{sec:overview}, we summarize and rederive the ACE algorithm 
using a superoperator notation, which differs from the original derivation in Ref.~\cite{ACE}
to clearly reveal the connection between the full environment Liouville 
space and the inner bonds of ACE PT-MPOs before MPO compression.
In Section~\ref{sec:derivation}, we explicitly derive the transformation 
matrices corresponding to the compression of inner indices during the ACE 
algorithm. We then discuss the resulting overall transformation matrices 
as well as their pseudoinverses in Section~\ref{sec:overall_trafo},
before describing how their explicit storage can be avoided by tracking
the transformation of observable closures in Section~\ref{sec:closures}.
The above-mentioned series of examples is presented in Section~\ref{sec:examples}.
Finally, our findings are summarized and discussed in Section~\ref{sec:discussion}.

\section{\label{sec:overview}Theoretical Background}
We first introduce the Liouville space notation used throughout the paper 
and summarize the basic concepts of influence functionals,
PT-MPOs, and the ACE algorithm.

\subsection{\label{sec:notation}Notation}
The time evolution of a density matrix of a general closed quantum system 
can be conveniently expressed in the superoperator formalism. 
To this end, we utilize the isomorphism between the Hilbert space 
$\mathcal{H}$, which hosts ``ket'' states $|\nu\rangle$, 
and its dual $\mathcal{H}^*$, which contains ``bra'' states $\langle \nu|$.
This enables the mapping of density matrices in the space
$\mathcal{H}\otimes \mathcal{H}^*$ onto vectors in the squared Hilbert space
$\mathcal{H}\otimes \mathcal{H}$, which we refer to as the Liouville space, by 
\begin{align}
&\hat{\rho}=\sum_{\nu,\mu=0}^{\textrm{dim}(\mathcal{H})-1}
\rho_{\nu\mu}|\nu\rangle\langle \mu| 
\nonumber\\&
\longleftrightarrow
|\rho)=\sum_{\nu,\mu=0}^{\textrm{dim}(\mathcal{H})-1} \rho_{\nu\mu} |\nu\rangle \otimes |\mu\rangle
=\sum_{\alpha=0}^{\left(\textrm{dim}(\mathcal{H})\right)^2-1} \rho_\alpha |\alpha),
\label{eq:Liouville_map}
\end{align}
where we have introduced combined indices $\alpha=(\nu,\mu)$ and 
defined the set of basis vectors $|\alpha)=|\nu\rangle\otimes |\mu\rangle$
of the Liouville space $\mathcal{H}\otimes \mathcal{H}$.
Thereby, the Liouville-von Neumann equation takes the form of a matrix-vector
product
\begin{align}
&\frac{\partial}{\partial t}\hat{\rho}= 
-\frac i\hbar\big[H,\hat{\rho}\big] 
=-\frac i\hbar\big(H\hat{\rho} - \hat{\rho}H\big)
\nonumber\\&
\longleftrightarrow \frac{\partial}{\partial t} |\rho) = 
-\frac i\hbar\big( H \otimes \mathbb{1} - \mathbb{1}\otimes H^T\big) |\rho)
=\mathcal{L}|\rho)
\label{eq:Liouville_closed}
\end{align}
with formal solution
\begin{align}
\label{eq:exp_formal_closed}
|\rho(t) )= e^{\mathcal{L} t} |\rho(0) ).
\end{align}
The inclusion of additional Markovian loss or decoherence processes via Lindblad terms and extension to a time-dependent Hamiltonian is straightforward.

We now consider a general open quantum system, where the composite system
$\mathcal{H}=\mathcal{H}_S \otimes \mathcal{H}_E$ 
can be decomposed into system $\mathcal{H}_S$ and environment subspaces
$\mathcal{H}_E$, respectively.
Henceforth, system Hilbert space basis states will be denoted by 
$|\nu\rangle$ and $|\mu\rangle$, while $|\xi\rangle$ and $|\eta\rangle$
will refer to basis states of the environment Hilbert space.
Instead of applying the superoperator mapping in Eq.~\eqref{eq:Liouville_map} 
directly to the total Hilbert space $\mathcal{H}$, we choose to first
rearrange indices in such a way that system $\alpha=(\nu,\mu)$ and environment 
degrees of freedom $\beta=(\xi,\eta)$ remain separate from each other:
\begin{align}
&\hat{\rho}=\sum_{\nu,\mu}\sum_{\xi,\eta} \rho_{\nu,\xi,\mu,\eta}
(|\nu\rangle\otimes|\xi\rangle) (\langle \mu|\otimes \langle \eta|)
\nonumber\\&
\longleftrightarrow
|\rho)=\sum_{\nu,\mu}\sum_{\xi,\eta} \rho_{(\nu,\mu),(\xi,\eta)}
|\nu\rangle\otimes|\mu\rangle\otimes |\xi\rangle \otimes |\eta\rangle 
\nonumber\\&
\quad\quad\;\;\,
=\sum_{\alpha,\beta} \rho_{\alpha,\beta} |\alpha)\otimes |\beta)
=\sum_{\alpha,\beta} \rho_{\alpha,\beta} |\alpha,\beta).
\label{eq:Liouville_map2}
\end{align}
In the final line of the above equation we have defined the components of the total density matrix 
$\rho_{\alpha,\beta}=(\alpha,\beta|\rho)$, where $(\alpha,\beta|=(\alpha|\otimes(\beta|$ with
$(\alpha|=\langle \nu|\otimes\langle\mu|$ and
$(\beta|=\langle \xi|\otimes|\langle \eta|$.
The time-evolution of the total density matrix is then formally given by 
integrating the Liouville-von Neumann equation
\begin{align}
\label{eq:Liouville_formal}
\rho_{\alpha,\beta}(t)=
& \sum_{\alpha_0,\beta_0}
(\alpha,\beta| e^{\mathcal{L} t}|\alpha_0,\beta_0)
%(\alpha_0,\beta_0|\rho(0)).
\rho_{\alpha_0,\beta_0}(0),
\end{align}
where $\rho_{\alpha_0, \beta_0}(0)$ are the coefficients of the joint initial state.
To extract a physical observable characterized by the operator
$\hat{O}$ acting on the total Hilbert space $\mathcal{H}$, we define 
\begin{align}
(\hat{O}|=& \sum_{\alpha,\beta} 
o_{\alpha,\beta} (\alpha,\beta|,
\\
\label{eq:Osupdef}
o_{(\nu,\mu),(\xi,\eta)} =& \big(\langle \mu | \otimes \langle \eta|\big)
\hat{O}\big(|\nu\rangle\otimes |\xi\rangle\big),
\end{align}
so that
\begin{align}
\label{eq:Osupav}
(\hat{O}| \rho)  =& \sum_{\alpha,\beta} o_{\alpha,\beta} \rho_{\alpha,\beta}
=\textrm{Tr}\big\{ \hat{O} \hat{\rho} \} =\langle \hat{O}\rangle.
\end{align}
Specifically, system observables $\hat{O}_S$ acting only on $\mathcal{H}_S$ 
are obtained by 
\begin{align}
\langle \hat{O}_S\rangle= (\hat{O}_S \otimes \mathbb{1}_E| \rho)=
\sum_{\alpha,\beta} o_\alpha \mathfrak{I}_\beta \rho_{\alpha,\beta}
=\sum_\alpha o_\alpha \bar{\rho}_{\alpha},
\label{eq:OS_Ifrak}
\end{align}
where $\bar{\rho}_{\alpha}=\sum_\beta \mathfrak{I}_\beta \rho_{\alpha,\beta}$
is the reduced system density matrix, 
$o_{\alpha}=o_{(\nu,\mu)}=\langle\mu|\hat{O}_S|\nu\rangle$, and 
$\mathfrak{I}_\beta=\mathfrak{I}_{(\xi,\eta)}=\delta_{\xi,\eta}$ denotes
the trace operation on the environment subspace.
Analogously, we define 
$\mathfrak{I}_\alpha=\mathfrak{I}_{(\nu,\mu)}=\delta_{\nu,\mu}$ to describe
the trace over the system degrees of freedom.

\subsection{\label{sec:FV}Influence functionals}
Our goal is to integrate the Liouville-von Neumann equation~\eqref{eq:Liouville_formal}
without explicitly operating on the full environment Liouville space because 
this is generally numerically infeasible. Instead, an exact
representation of the effects of the environment can be formulated, 
where the explicit environment degrees of freedom are traced out. 
The corresponding object is the \emph{influence functional}, which was first
derived by Feynman and Vernon in the context of 
real-time path integrals~\cite{FeynmanVernon}. 

However, we shall now instead consider an alternative derivation within the superoperator formalism that is obtained in three steps:
First, time is discretized on a regular grid 
$t_j = t_0 + j \Delta t$ with time steps $\Delta t$, yielding the time evolution
\begin{align}
&(\alpha_n,\beta_n| e^{\mathcal{L} t_n}|\alpha_0,\beta_0) \nonumber\\
&=\sum_{\substack{\alpha_{n-1},\dots,\alpha_{1} \\\beta_{n-1},\dots,\beta_{1}}}
\prod_{l=1}^{n} 
(\alpha_l,\beta_l| e^{\mathcal{L} \Delta t}|\alpha_{l-1},\beta_{l-1}).
\end{align}
To simplify notation, time arguments are henceforth implied in the
index labels. For example, the subscript $j$ on the system Liouville space index
$\alpha_j$ indicates that 
$\bar{\rho}_{\alpha_j}$ denotes $\bar{\rho}_{\alpha_j}(t_j)$,
i.e.~the reduced system density matrix at time $t_j$.

Second, the total Liouvillian is decomposed into
$\mathcal{L}=\mathcal{L}_S+\mathcal{L}_E$,
where the system Liouvillian $\mathcal{L}_S$ only affects the system, while 
the environment Liouvillian $\mathcal{L}_E$ includes the system-environment 
interaction and, hence, affects both system and environment. 
The time evolution within each time step is then split using 
the Trotter decomposition
\begin{align}
e^{(\mathcal{L}_E+\mathcal{L}_S)\Delta t} =
e^{\mathcal{L}_E\Delta t}e^{\mathcal{L}_S\Delta t} +\mathcal{O}(\Delta t^2).
\label{eq:Trotter}
\end{align}
The implementation of a symmetric Trotter decomposition with error 
$\mathcal{O}(\Delta t^3)$ is straightforward, but we proceed our derivation
with Eq.~\eqref{eq:Trotter} for a shorter notation.

Finally, assuming that the initial state 
$\rho_{\alpha_0,\beta_0}=\bar{\rho}_{\alpha_0}\rho^{E}_{\beta_0}$ 
factorizes into system $\bar{\rho}_{\alpha_0}$ and environment parts
$\rho^{E}_{\beta_0}$, one traces over the environment at the final time step.
Then, the reduced system density matrix at time $t=t_n$ can be expressed as
\begin{align}
\bar{\rho}_{\alpha_n}=&\sum_{\substack{
\alpha_{n-1},\ldots,\alpha_0 \\
\alpha'_n,\ldots,{\alpha}'_1}}
\mathcal{I}^{(\alpha_n,{\alpha}'_n)\ldots (\alpha_1,{\alpha}'_1)}
\bigg(\prod_{l=1}^n
\mathcal{M}^{\alpha'_l \alpha_{l-1}} \bigg)
\bar{\rho}_{\alpha_0},
\label{eq:rho_alpha}
\end{align}
where $\mathcal{M}^{{\alpha}'_l \alpha_{l-1}}= 
(\alpha'_l|e^{\mathcal{L}_S \Delta t}| \alpha_{l-1})$
denotes the free system propagator and
\begin{align}
&\mathcal{I}^{(\alpha_n,\alpha'_n)\ldots (\alpha_1,\alpha'_1)}
\nonumber\\
&=\sum_{\beta_{n},\ldots,\beta_1} \mathfrak{I}_{\beta_{n}}
\bigg(\prod_{l=1}^{n} 
(\alpha_l,\beta_l|e^{\mathcal{L}_E \Delta t}|\alpha'_l,\beta_{l-1})\bigg)
\rho^E_{\beta_0}
\label{eq:IFlong}
\end{align}
is the influence functional. 
The sum over all possible combinations of system indices $\alpha_l$ and 
$\alpha'_l$ in Eq.~\eqref{eq:rho_alpha} becomes the integral over all
system paths in the continuous-time limit described by the 
Feynman-Vernon real-time path integral formalism~\cite{FeynmanVernon}.

If the environment is Gaussian, as in the case of the spin-boson model,
explicit expressions for the influence functional can be obtained by 
analytically integrating over the environment degrees of freedom 
$\beta_{l}$~\cite{FeynmanVernon}, which is used in various algorithms including QUAPI~\cite{QUAPI1}, TEMPO~\cite{TEMPO}, and others~\cite{JP,DnC,Link_infinite}.
A numerical approach for more general environments is provided by ACE~\cite{ACE}, where environment influences are calculated explicitly using Eq.~\eqref{eq:IFlong}. A key element to make ACE numerically feasible is the MPO representation of the environment influences, which we discuss next.

\subsection{\label{sec:PT}Process tensor matrix product operators}
The Feynman-Vernon path sum in Eq.~\eqref{eq:rho_alpha}
yields an exact description of the open quantum systems dynamics in the limit
$\Delta t\to 0$.
However, its practical application is limited by the exponential
scaling of the number of paths $\mathcal{O}(\textrm{dim}(\mathcal{H_S})^{4n})$
with the number of time steps $n$.
A practical solution is provided by PT-MPOs~\cite{PT_PRA,JP}, which represent
influence functionals efficiently in the form of 
MPOs~\cite{MPS_Orus,MPS_Schollwoeck}
\begin{align}
&\mathcal{I}^{(\alpha_n,\alpha'_n)\ldots (\alpha_1,\alpha'_1)}
\nonumber\\ &\quad\quad\;=
\sum_{d_{n},\dots,d_0}
\mathcal{Q}^{(\alpha_{n},\alpha'_{n})}_{d_n d_{n-1}}
\mathcal{Q}^{(\alpha_{n-1},\alpha'_{n-1})}_{d_{n-1} d_{n-2}}
\ldots
\mathcal{Q}^{(\alpha_1,\alpha'_1)}_{d_1 d_0}.
\label{eq:IfromQ}
\end{align}
Here, $\mathcal{Q}^{(\alpha_{l},\alpha'_{l})}_{d_l d_{l-1}}$ are interpreted
as matrices with respect to inner bond indices $d_l$, where
bond indices at the edges take only one value $d_0=d_n=0$. 
The MPO form makes it possible to perform the path summation in
Eq.~\eqref{eq:rho_alpha} step by step.
To this end, we define the extended density matrix $\rho^{\alpha}_d$ by the iteration
\begin{subequations}
\label{eq:iter}
\begin{align}\rho^{\alpha_0}_{0}=&\bar{\rho}_{\alpha_0},
\\
\rho^{\alpha_l}_{d_l}=&\sum_{\alpha'_l,\alpha_{l-1},d_{l-1}}
\mathcal{Q}^{(\alpha_{l},\alpha'_{l})}_{d_l d_{l-1}}
\mathcal{M}^{\alpha'_l \alpha_{l-1}}  \rho^{\alpha_{l-1}}_{d_{l-1}},
\end{align}
\end{subequations}
which turns the sum over exponentially many paths into a linear number of
matrix multiplications. The reduced system density matrix at the last time step $t_n$
is then given by $\bar{\rho}_{\alpha_n}=\rho^{\alpha_n}_{0}$. At intermediate time steps it can be obtained by 
$\bar{\rho}_{\alpha_l}=\sum_{d_l} q_{d_l}\rho^{\alpha_l}_{d_l}$, 
where the closures $q_{d_l}$ are constructed from the PT-MPO as described in
Ref.~\cite{ACE}.
Thus, the brunt of the work required to simulate the open quantum system
is now shifted to bringing the influence functional into the PT-MPO form 
of Eq.~\eqref{eq:IfromQ}.

In principle, any tensor can be brought into MPO form 
by successive Schmidt decompositions or, equivalently,
singular value decompositions (SVDs). 
It is straightforward to show~\cite{MPS_Orus} that this provides an upper 
bound for the inner bond dimensions by 
$d_j,d_{n-j}\le \textrm{dim}(\mathcal{H}_S^{4 j})$, which is maximal
at the center of the chain, e.g., 
$d_{n/2}\le \textrm{dim}(\mathcal{H}_S^{2n})$ for even $n$. 
However, often, many of the singular values are zero or negligibly small,
which reduces the inner bond dimensions significantly.
Furthermore, to avoid exponential scaling incurred by the factorization of
the full tensor, PT-MPOs are usually built up from smaller 
blocks, keeping them in MPO form at all times~\cite{JP, ACE, DnC}.
After adding new blocks, the PT-MPO is compressed by sweeping along the MPO 
chain and reducing inner dimensions using truncated SVDs, 
where singular values below a given threshold are 
neglected~\cite{MPS_Schollwoeck}.
For example, converged results have been obtained 
in typical applications with compressed PT-MPOs 
with maximal inner dimensions $\chi$ in the range of a few dozen 
to several hundreds~\cite{ACE,DnC}.

\subsection{Automated compression of environments}

The ACE algorithm~\cite{ACE} enables the calculation of PT-MPOs also
for non-Gaussian environments, for which no analytical expressions based
on path integration are available. Instead, Eq.~\eqref{eq:IFlong}
is directly brought into the MPO form of Eq.~\eqref{eq:IfromQ} starting 
from the microscopic Hamiltonian. ACE is based on the observation that,
in our Liouville space notation, an exact PT-MPO can be formally constructed 
by setting
\begin{subequations}
\label{eq:formalPTMPO}
\begin{align}
\mathcal{Q}^{(\alpha_l,\alpha'_l)}_{\beta_{l},\beta_{l-1}}=
(\alpha_l, \beta_l| e^{\mathcal{L}_E \Delta t} |\alpha'_l,\beta_{l-1})
\end{align}
for matrices inside the MPO chain $2\le l\le n-1$ and
\begin{align}
\mathcal{Q}^{(\alpha_1,\alpha'_1)}_{\beta_{1},\beta_{0}}=
\sum\limits_{\beta'}
(\alpha_1, \beta_1| e^{\mathcal{L}_E \Delta t} |\alpha'_1,\beta')
\rho^E_{\beta'} \, \delta_{\beta_{0},0}
\end{align}
and
\begin{align}
\mathcal{Q}^{(\alpha_n,\alpha'_n)}_{\beta_{n},\beta_{n-1}}=
\delta_{\beta_{n},0}\sum\limits_{\beta'} \mathfrak{I}_{\beta'}
(\alpha_n, \beta'| e^{\mathcal{L}_E \Delta t} |\alpha'_n,\beta_{n-1}),
\end{align}
\end{subequations}
for the first and the last MPO matrix, respectively.

Note, however, that the inner bond dimension of this PT-MPO is equal to the
dimension of the full environment Liouville space  
$\textrm{dim}(\mathcal{H}_E)^2$. Hence, the propagatation using the iteration
Eq.~\eqref{eq:iter} is as computationally expensive as solving the time
evolution of the total system comprised of system of interest and environment
in Liouville space.

ACE overcomes this challenge by considering the environment as being composed
of $N_E$ independent degrees of freedom, henceforth referred to as 
\textit{modes}. 
The total environmental Hilbert space is then a product of individual subspaces
$\mathcal{H}_E=\mathcal{H}_E^{(1)}\otimes\mathcal{H}_E^{(2)}\otimes
\dots\mathcal{H}_E^{(N_E)}$, whilst the 
environment Hamiltonian is given by the sum
$H_E = \sum_{k=1}^{N_E} H_E^{(k)}$,
where each summand $H_E^{(k)}$ only operates on the system and the $k$-th environment mode 
Hilbert space $\mathcal{H}_S\otimes \mathcal{H}_E^{(k)}$. Consequently, the total Liouvillian can also be written as a sum of $N_E$ terms
$\mathcal{L}_E=\sum_k^{N_E}\mathcal{L}_E^{(k)}$, where the $k$-th environment
Liouvillian
has the explicit matrix representation
\begin{align}
&(\alpha,\beta^{(k)}|\mathcal{L}_E^{(k)}|\alpha',{\beta'}^{(k)})\nonumber\\
&=((\nu,\mu),(\xi^{(k)},\eta^{(k)})|\mathcal{L}_E^{(k)}|(\nu',\mu'),({\xi'}^{(k)},{\eta'}^{(k)}))\nonumber\\
&=-\frac i\hbar \Big[ \langle \nu, \xi^{(k)}|H_E^{(k)}|\nu',{\xi'}^{(k)}\rangle \delta_{\mu,\mu'} \delta_{\eta^{(k)},{\eta'}^{(k)}}\nonumber\\
&\qquad-\langle \mu', {\eta'}^{(k)}|H_E^{(k)}|\mu,{\eta}^{(k)}\rangle \delta_{\nu,\nu'} \delta_{\xi^{(k)},{\xi'}^{(k)}}
\Big],
\label{eq:LkMatrix}
\end{align}
where indices $\beta^{(k)}=(\xi^{(k)},\eta^{(k)})$ enumerate a basis of states for the $k$-th environment Liouville space.
%the matrix representation of 
The $k$-th environment propagator 
$(\alpha,\beta^{(k)}|e^{\mathcal{L}_E^{(k)}\Delta t}|\alpha',{\beta'}^{(k)})$ is then given by the matrix exponential of Eq.~\eqref{eq:LkMatrix}.

For each individual environment mode $k$, a PT-MPO is constructed
using Eqs.~\eqref{eq:formalPTMPO} with the Liouvillian $\mathcal{L}_E$ replaced
by $\mathcal{L}_E^{(k)}$. 
It is assumed that the inner bond dimension, which now corresponds to
the Liouville space of only a single environment mode, is manageable.
The PT for the full environment is obtained by combining the PTs for all 
individual modes based on the sequential application of the
symmetric Trotter decomposition
\begin{align}
e^{\sum\limits_{k=1}^{K}\mathcal{L}_E^{(k)}\Delta t} \approx
e^{\mathcal{L}_E^{(K)}\tfrac{\Delta t}2} 
\bigg(e^{\sum\limits_{k=1}^{K-1}\mathcal{L}_E^{(k)} \Delta t} \bigg)
e^{\mathcal{L}_E^{(K)}\tfrac{\Delta t}2} 
\label{eq:symTrotterK}
\end{align}
for $K=2,\dots N_E$, which incurs a Trotter error $\mathcal{O}(\Delta t^3)$.

To keep the bond dimensions reasonably small, the PT-MPO is compressed after 
the inclusion of each additional environment mode, as described in detail in the 
next section.
Eventually, after having iteratively incorporated the PT-MPO of each environmental mode, 
one arrives at a PT-MPO describing all effects of the microscopic
Hamiltonian numerically exactly, where the only numerical errors are due
to the Trotter decomposition and the MPO compression.

\section{Transformation of inner bonds of process tensors\label{sec:derivation}}
Our goal now is to explicitly link the inner bonds of PT-MPOs
$\mathcal{Q}^{(\alpha_l,\alpha'_l)}_{d_l,d_{l-1}}$ 
to the environment propagator in Liouville space.
Concretely, we show that the PT-MPO matrices obtained using ACE can
be written as
\begin{align}
\mathcal{Q}^{(\alpha_l,\alpha'_l)}_{d_l,d_{l-1}}=&
\sum_{\beta_l,\beta_{l-1}} \mathcal{T}_{d_l,\beta_l}
(\alpha_l,\beta_l|e^{\mathcal{L}_E\Delta t}|\alpha'_l,\beta_{l-1})
\mathcal{T}^{-1}_{\beta_{l-1}, d_{l-1}},
\label{eq:QfromT}
\end{align}
where $\mathcal{T}_{d_l,\beta_l}$ describes a lossy transformation and
$\mathcal{T}^{-1}_{\beta_{l-1}, d_{l-1}}$ is a pseudoinverse
of the corresponding transformation matrix $\mathcal{T}_{d_{l-1},\beta_{l-1}}$
at the previous time step $t_{l-1}$.
To this end, we follow the evolution of the transformation matices 
$\mathcal{T}_{d_l,\beta_l}$ and their pseudoinverses in every step
of the ACE algorithm.

As a corollary of Eq.~\eqref{eq:QfromT}, 
we observe that the propagated quantity $\rho^{\alpha_l}_{d_l}$ 
in Eq.~\eqref{eq:iter} corresponds to 
the total density matrix after compression of the inner bonds
$\rho^{\alpha_l}_{d_l}=
\sum_{\beta_l} \mathcal{T}_{d_l,\beta_l}\rho_{\alpha_l,\beta_l}$,
which can be written in the notation of
section~\ref{sec:notation} as $|\mathcal{T} \rho(t) )$.
This suggests that general observables $\hat{O}$, including environment and 
mixed system-environment observables, can be inferred from
\begin{align}
\langle \hat{O}(t)\rangle\approx(\hat{O} \mathcal{T}^{-1}|
\mathcal{T} \rho(t)),
\label{eq:Oextract_formal}
\end{align}
which, however, only holds approximately due to the lossy nature of the
transformation $\mathcal{T}$. 
Its validity is tested on several examples in Section~\ref{sec:examples}.

The advantage of Eq.~\eqref{eq:Oextract_formal} is that it provides an 
efficient way to extract environment observables without the need store
the lossy transformation matrices $\mathcal{T}$ and $\mathcal{T}^{-1}$
explicitly. It suffices to track only the behaviour of 
$(\hat{O}\mathcal{T}^{-1}|$ for given observables $\hat{O}$.

\subsection{PT-MPO combination}
We now describe how the transformations $\mathcal{T}$ and 
$\mathcal{T}^{-1}$ in Eq.~\eqref{eq:QfromT} are affected 
in the individual combination and compression
steps of the ACE algorithm~\cite{ACE}. We do this by induction: 
We start with the trivial PT-MPO with matrices 
$\mathcal{Q}^{(\alpha_l,\alpha'_l)}_{d_l,d_{l-1}}=
\delta_{\alpha_l,\alpha'_l}\delta_{d_l,0}\delta_{d_{l-1},0}$, which correspond to a dummy environment mode of dimension dim($\mathcal{H}_E^{(0)}$)=1 with Hamiltonian $H_E^{(0)}=0$. The initial transformation matrices are  
$\mathcal{T}_{d_l,\beta_l}=\mathcal{T}^{-1}_{\beta_l,d_l}=
\delta_{d_l,0}\delta_{\beta_l,0}$.

In the induction step, it has to be shown that, if the MPO matrices
of the PT accounting for modes $k=1,\dots, K-1$ 
are of the form in Eq.~\eqref{eq:QfromT}, then the resulting MPO matrices
of the PT including the influence of the mode $K$ are also
of the form in Eq.~\eqref{eq:QfromT}, and the respective transformation matrices
are related by well-defined linear operations. 

Denoting MPO matrices
of the input PT-MPO by $\mathcal{Q}^{(\alpha_l,\alpha'_l)}_{d_l,d_{l-1}}$ and
that of the resulting PT-MPO by 
$\tilde{\mathcal{Q}}^{(\tilde{\alpha}_l,\tilde{\alpha}'_l)}_{\tilde{d}_l,\tilde{d}_{l-1}}$,
it follows from the symmetric Trotter formula in Eq.~\eqref{eq:symTrotterK}
that
\begin{align}
\tilde{\mathcal{Q}}^{(\tilde{\alpha}_l,\tilde{\alpha}'_l)}_{(d_l,\beta_l),(d_{l-1},\beta_{l-1})}\approx
\sum_{\alpha_l,\alpha'_l,\beta'}
\mathcal{B}^{(\tilde{\alpha}_l,\alpha_l)}_{\beta_{l},\beta'}
\mathcal{Q}^{(\alpha_l,\alpha'_l)}_{d_l,d_{l-1}}
\mathcal{B}^{(\alpha'_l,\tilde{\alpha}'_l)}_{\beta',\beta_{l-1}}
\label{eq:BQB}
\end{align}
with matrix representation of the $K$-th environment Liouville propagator 
for half a time step
$\mathcal{B}^{(\tilde{\alpha}_l,\alpha_l)}_{\beta_{l},\beta'} =
(\tilde{\alpha}_l,\beta_l|e^{\mathcal{L}_E^{K}\tfrac{\Delta t}2}|
\alpha_l,\beta')$.
The relation to Eq.~\eqref{eq:QfromT} is established by identifying
\begin{widetext}
\begin{align}
\label{eq:induction}
&\mathcal{Q}^{(\alpha_l,\alpha'_l)}_{d_l,d_{l-1}}\approx 
\sum_{\substack{\beta_l^{(1)},\dots,\beta_l^{(K-1)}\\
\beta_{l-1}^{(1)},\dots,\beta_{l-1}^{(K-1)}}}
\mathcal{T}_{d_l,(\beta_l^{(1)}, \dots, \beta_l^{(K-1)})}
(\alpha_l,\beta_l^{(1)},\dots,\beta_l^{(K-1)}|
e^{\sum\limits_{k=1}^{K-1} \mathcal{L}_E^{(k)} \Delta t}
|\alpha'_l,\beta_{l-1}^{(1)},\dots,\beta_{l-1}^{(K-1)}) 
\mathcal{T}^{-1}_{(\beta_{l-1}^{(1)},\dots,\beta_{l-1}^{(K-1)}), d_{l-1}}
\nonumber\\
&\Longrightarrow
\tilde{\mathcal{Q}}^{(\tilde{\alpha}_l,\tilde{\alpha}'_l)}_{\tilde{d}_l,\tilde{d}_{l-1}}\approx 
\sum_{\substack{\beta_l^{(1)},\dots,\beta_l^{(K)}\\
\beta_{l-1}^{(1)},\dots,\beta_{l-1}^{(K)}}}
\tilde{\mathcal{T}}_{\tilde{d}_l,(\beta_l^{(1)}, \dots, \beta_l^{(K)})}
(\tilde{\alpha}_l,\beta_l^{(1)},\dots,\beta_l^{(K)}|
e^{\sum\limits_{k=1}^{K} \mathcal{L}_E^{(k)} \Delta t}
|\tilde{\alpha}'_l,\beta_{l-1}^{(1)},\dots,\beta_{l-1}^{(K)}) 
\tilde{\mathcal{T}}^{-1}_{(\beta_{l-1}^{(1)},\dots,\beta_{l-1}^{(K)}), \tilde{d}_{l-1}}
\end{align}
\end{widetext}
where the original transformation matrices $\mathcal{T}$ and $\mathcal{T}^{-1}$
are changed into  
\begin{subequations}
\label{eq:Texpand}
\begin{align}
\tilde{\mathcal{T}}_{\tilde{d}_l,(\beta_l^{(1)}, \dots, \beta_l^{(K)})}
=&\delta_{\tilde{d}_l, (d_l,\beta_l^{(K)})}
\mathcal{T}_{d_l,(\beta_l^{(1)}, \dots, \beta_l^{(K-1)})},
\\
\label{eq:Texpand_b}
\tilde{\mathcal{T}}^{-1}_{(\beta_l^{(1)}, \dots, \beta_l^{(K)}),\tilde{d}_l}
=&\mathcal{T}^{-1}_{(\beta_l^{(1)}, \dots, \beta_l^{(K-1)}),d_l}
\delta_{(d_l,\beta_l^{(K)}),\tilde{d}_l},
\end{align}
\end{subequations}
such that the original inner bonds $d_l$ are simply expanded
to include the Liouville space of the $K$-th environment mode 
described by basis states with indices $\beta_l^{(K)}$.

\subsection{MPO compression}
Combining PT-MPOs of individual environment modes increases the 
inner dimension of the resulting PT-MPO. To keep the combined PT-MPO tractable, it undergoes a compression step 
after every combination. Compression is achieved by sweeping
along the MPO chain, first forward, then backward, while reducing the inner 
dimension by truncated SVDs~\cite{JP,ACE}. 

SVDs decompose arbitrary $n\times m$ matrices $A$ 
\begin{align}
A=U \Sigma V^\dagger,
\label{eq:SVD}
\end{align}
where $U$ and $V$ are matrices with orthogonal columns, while 
$\Sigma$ %$=\textrm{diag}(\sigma_0,\sigma_1,\dots,\sigma_s)$ 
is a diagonal matrix
with the non-negative real singular values $\sigma_j$ on the diagonal, 
which are assumed to be sorted in descending order, 
i.e. $\sigma_j\ge \sigma_{k}$ for $j\le k$.
Here, we use truncated SVDs by keeping only the $s$ most significant 
singular values $\sigma_j\ge \epsilon\sigma_0$, 
where $\epsilon$ is a given truncation threshold relative 
to the largest singular value $\sigma_0$, and 
also restricting the matrices $U$ and $V$ to their first $s$ columns.
For our applications, $s$ will be the inner bond dimension after PT-MPO 
compression. 

In fact, the Eckart-Young-Mirsky theorem~\cite{eckart1936} states that
SVDs provide the best low-rank approximation to a general matrix for given
rank $s$.
However, MPO compression at a given site using truncated SVD is only locally optimal if the MPO is in mixed canonical form~\cite{MPS_Schollwoeck}. One way to achieve this is to first perform a forward sweep with non-truncating SVDs and only
truncating during the backward sweep. Yet local optimal compression is not necessarily needed for the most efficient algorithm. Here, we use a strategy in the spirit of the zip-up algorithm of Ref.~\cite{zipup}, where truncation is performed for both, forward and backward sweeps. The sizable speed-up achieved by operating on matrices with smaller dimensions is typically well worth the slight increase in numerical error, especially because the latter can be mitigated by using a smaller compression threshold. As discussed in Appendix A of Ref.~\cite{DnC} as well as in Ref.~\cite{combine_tree}, choosing different compression parameters for different sweeps can be used to fine-tune PT-MPO algorithms. For example, using a different threshold $\epsilon_{fw}$ for forward sweeps compared to backward sweeps $\epsilon_{bw}$, one can interpolate between the locally optimal compression ($\epsilon_{fw}=0$) and the zip-up-like approach ($\epsilon_{fw}=\epsilon_{bw}$). Here, we use $\epsilon_{fw}=\epsilon_{bw}=\epsilon$ throughout this article.
Furthermore, note that MPO compression by SVDs in general is aimed at minimizing the norm distance between MPOs before and after compression. To our knowledge, no formal theory linking this norm distance to a precise error bound for physical observables is available so far. However, numerical simulations in Ref.~\cite{combine_tree} indicate that the PT-MPO norm distance and system observables show similar convergence behavior. This provides empirical justification for PT-MPO compression using sweeps with SVDs as described above.

Next, we summarize how MPO matrices and in particular their inner bonds are affected by forward and backward sweeps.
To simplify the notation, we henceforth imply truncation
whenever we refer to SVDs like in Eq.~\eqref{eq:SVD}.

\subsubsection{Forward sweep}
First, we compress the PT-MPO by sweeping along the MPO in the forward direction,
i.e., from $l=1$ to $l=n$. At step $l$, we perform an SVD 
on the PT-MPO matrices $\mathcal{Q}^{(\alpha_l,\alpha'_{l})}_{d_l,d_{l-1}}$,
which we interpret as a single matrix 
$A_{d_l, ((\alpha_l,\alpha'_{l}),d_{l-1})}$ with outer indices combined
with the right inner index $d_{l-1}$.
\begin{align}
\mathcal{Q}^{(\alpha_l,\alpha'_{l})}_{d_l,d_{l-1}}
=\sum_{j_l} 
\overleftarrow{U}_{d_l j_l} \overleftarrow{\sigma}_{j_l}
{\overleftarrow{V}}^{\dagger}_{j_l ((\alpha_l,\alpha'_l),d_{l-1})}.
\label{eq:forwardSVD}
\end{align}
After truncation, the MPO matrix at steps $l$ is replaced by
\begin{subequations}
\begin{align}
\tilde{\mathcal{Q}}^{(\alpha_l,\alpha'_{l})}_{j_l,d_{l-1}}
={\overleftarrow{V}}^{\dagger}_{j_l ((\alpha_l,\alpha'_l),d_{l-1})},
\end{align}
while $\overleftarrow{U}_{d_l j_l}$ and the singular values 
$\overleftarrow{\sigma}_{j_l}$ are passed on to the next MPO matrix at
step $l+1$ 
\begin{align}
\tilde{\mathcal{Q}}^{(\alpha_{l+1},\alpha'_{l+1})}_{d_{l+1},j_l}
=\sum_{d_l}\mathcal{Q}^{(\alpha_{l+1},\alpha'_{l+1})}_{d_{l+1},d_{l}}
\overleftarrow{U}_{d_l j_l} \overleftarrow{\sigma}_{j_l}.
\end{align}
\end{subequations}
Passing on the singular values $\overleftarrow{\sigma}_{j_l}$,
which are indicators of the local importance of the respective 
indices $j_l$, is crucial for efficient compression aimed at selecting
degrees of freedom that are relevant for describing the system dynamics 
over many time steps.

Alternatively, the updated MPO matrices can be expressed as
\begin{subequations}
\begin{align}
\tilde{\mathcal{Q}}^{(\alpha_l,\alpha'_{l})}_{j_l,d_{l-1}}
=\sum_{d_l}\overleftarrow{T}_{j_l, d_{l}} 
\mathcal{Q}^{(\alpha_l,\alpha'_{l})}_{d_l,d_{l-1}},\\
\tilde{\mathcal{Q}}^{(\alpha_{l+1},\alpha'_{l+1})}_{d_{l+1},j_l}
=\sum_{d_l}\mathcal{Q}^{(\alpha_{l+1},\alpha'_{l+1})}_{d_{l+1},d_{l}} 
\overleftarrow{T}^{-1}_{d_l, j_l}
\end{align}
\end{subequations}
with transformation matrices
\begin{subequations}
\begin{align}
\overleftarrow{T}_{j_l, d_{l}}
=& \overleftarrow{\sigma}_{j_l}^{-1}
\overleftarrow{U}^\dagger_{j_l d_{l}},
\\
\overleftarrow{T}^{-1}_{d_l, j_l}
=&\overleftarrow{U}_{d_l j_l} \overleftarrow{\sigma}_{j_l}.
\end{align}
\label{eq:T_for}
\end{subequations}
Assuming that the original PT-MPO matrices have been expressible in terms
of environment propagators after lossy compression as in
Eq.~\eqref{eq:QfromT},
the corresponding transformation matrices $\mathcal{T}$ and $\mathcal{T}^{-1}$ 
are themselves transformed as
\begin{subequations}
\label{eq:Ttrafo_for}
\begin{align}
\tilde{\mathcal{T}}_{j_l,\beta_l}=&\sum_{d_l}
\overleftarrow{T}_{j_l, d_{l}} \mathcal{T}_{d_l,\beta_l},
\\
\label{eq:Ttrafo_for_b}
\tilde{\mathcal{T}}^{-1}_{\beta_l,j_l}=&\sum_{d_l}
\mathcal{T}^{-1}_{\beta_l,d_l} \overleftarrow{T}^{-1}_{d_l, j_l}.
\end{align}
\end{subequations}

\subsubsection{Backward sweep}
Compression is enhanced, if the forward sweep over the PT-MPO is followed by 
a backward sweep from $l=n$ to $l=2$. 
Here, the outer indices of the MPO matrices 
$\mathcal{Q}^{(\alpha_l,\alpha'_{l})}_{d_l,d_{l-1}}$ are combined with the
left inner index $d_l$ to form the matrix 
$A_{(d_l,(\alpha_l,\alpha'_{l})),d_{l-1}}$, whose SVD yields
\begin{align}
\mathcal{Q}^{(\alpha_l,\alpha'_{l})}_{d_l,d_{l-1}}
=\sum_{j_{l-1}}
\overrightarrow{U}_{(d_l,(\alpha_l,\alpha'_{l})), j_{l-1}}
\overrightarrow{\sigma}_{j_{l-1}}
{\overrightarrow{V}}^{\dagger}_{j_{l-1},d_{l-1}}.
\end{align}
Now, the MPO matrix at step $l$ is updated as 
\begin{subequations}
\begin{align}
\tilde{\mathcal{Q}}^{(\alpha_l,\alpha'_{l})}_{d_l,j_{l-1}}
=\overrightarrow{U}_{(d_l,(\alpha_l,\alpha'_{l})), j_{l-1}}
\end{align}
and the singular values as well as the matrix 
${\overrightarrow{V}}^{\dagger}_{j_{l-1},d_{l-1}}$ are passed on the prior
step $l-1$
\begin{align}
\tilde{\mathcal{Q}}^{(\alpha_{l-1},\alpha'_{l-1})}_{j_{l-1}, d_{l-2}}
=\sum_{d_{l-1}}\overrightarrow{\sigma}_{j_{l-1}}
{\overrightarrow{V}}^{\dagger}_{j_{l-1},d_{l-1}}
\mathcal{Q}^{(\alpha_{l-1},\alpha'_{l-1})}_{d_{l-1}, d_{l-2}}.
\end{align}
\end{subequations}

Analogously to the forward sweep, we cast the update into the shape of a
transformation 
\begin{subequations}
\begin{align}
\tilde{\mathcal{Q}}^{(\alpha_l,\alpha'_{l})}_{d_l,j_{l-1}}
=\sum_{d_{l-1}} \mathcal{Q}^{(\alpha_l,\alpha'_{l})}_{d_l,d_{l-1}}
\overrightarrow{T}^{-1}_{d_{l-1}, j_{l-1}},
\\
\tilde{\mathcal{Q}}^{(\alpha_{l-1},\alpha'_{l-1})}_{j_{l-1},d_{l-2}}
=\sum_{d_{l-1}}\overrightarrow{T}_{j_{l-1}, d_{l-1}} 
\mathcal{Q}^{(\alpha_{l-1},\alpha'_{l-1})}_{d_{l-1},d_{l-2}}
\end{align}
\end{subequations}
with 
\begin{subequations}
\begin{align}
\overrightarrow{T}_{j_{l-1}, d_{l-1}}
=& \overrightarrow{\sigma}_{j_{l-1}}
\overrightarrow{V}^\dagger_{j_{l-1} d_{l-1}},
\\
\overrightarrow{T}^{-1}_{d_{l-1}, j_{l-1}}
=&\overrightarrow{V}_{d_{l-1} j_{l-1}} \overrightarrow{\sigma}^{-1}_{j_{l-1}}.
\end{align}
\label{eq:T_back}
\end{subequations}

The overall transformation matrices in Eq.~\eqref{eq:QfromT} are modified as
\begin{subequations}
\label{eq:Ttrafo_back}
\begin{align}
\tilde{\mathcal{T}}_{j_l,\beta_l}=&\sum_{d_l}
\overrightarrow{T}_{j_l, d_{l}} \mathcal{T}_{d_l,\beta_l},
\\
\label{eq:Ttrafo_back_b}
\tilde{\mathcal{T}}^{-1}_{\beta_l,j_l}=&\sum_{d_l}
\mathcal{T}^{-1}_{\beta_l,d_l} \overrightarrow{T}^{-1}_{d_l, j_l}.
\end{align}
\end{subequations}

\subsection{\label{sec:overall_trafo}Overall transformation}
We are now in the position to formulate an explicit expression for 
the transformation matrices in Eq.~\eqref{eq:QfromT}
relating the full environment propagator to the final PT-MPO matrices 
obtained by the ACE algorithm, which is visualized in Fig.~\ref{fig:sketch}(c).

To this end, we concatenate for each of the $N_E$ environment modes
the expansion step in Eqs.~\eqref{eq:Texpand} with the forward and backward 
sweeps in Eqs.~\eqref{eq:Ttrafo_for} and \eqref{eq:Ttrafo_back}, respectively.
Combining
\begin{subequations}
\begin{align}
\overleftrightarrow{T}_{d_l^{(k)},(d_{l}^{(k-1)},\beta^{(k)}_l)}=&
\sum_{j_l^{(k)}}
\overrightarrow{T}_{d_l^{(k)},j_l^{(k)}}
\overleftarrow{T}_{j_l^{(k)},(d_{l}^{(k-1)},\beta^{(k)}_l)},
\\
\overleftrightarrow{T}^{-1}_{(d_{l}^{(k-1)},\beta^{(k)}_l),d_l^{(k)}}=&
\sum_{j_l^{(k)}}
\overleftarrow{T}^{-1}_{(d_{l}^{(k-1)},\beta^{(k)}_l),j_l^{(k)}}
\overrightarrow{T}^{-1}_{j_l^{(k)},d_l^{(k)}},
\end{align}
\end{subequations}
the overall transformation matrices become
\begin{subequations}
\label{eq:Ttotal}
\begin{align}
\mathcal{T}_{d_l^{(N_E)}, (\beta_l^{(1)},\dots,\beta_l^{(N_E)})} =&
\sum_{d_l^{(1)},\dots,d_l^{(N_E-1)}}\prod_{k=1}^{N_E} 
\overleftrightarrow{T}_{d_l^{(k)},(d_{l}^{(k-1)},\beta^{(k)}_l)},
\\
\label{eq:Ttotal_b}
\mathcal{T}^{-1}_{(\beta_l^{(1)},\dots,\beta_l^{(N_E)}),d_l^{(N_E)}} =&
\sum_{d_l^{(1)},\dots,d_l^{(N_E-1)}} \prod_{k=1}^{N_E}
\overleftrightarrow{T}^{-1}_{(d_{l}^{(k-1)},\beta^{(k)}_l),d_l^{(k)}},
\end{align}
\end{subequations}
where the index $d_l^{(0)}$ is understood to be restricted to $d_l^{(0)}=0$, 
leading to no expansion for the double index 
$(d_{l}^{(0)},\beta^{(1)}_l)=\beta^{(1)}_l$.

Several aspects of Eqs.~\eqref{eq:Ttotal} are noteworthy:
First, it is instructive to compare the lossy transformation induced
by $\mathcal{T}$ and its pseudoinverse $\mathcal{T}^{-1}$ with a
conventional projection to a reduced subspace spanned by a set of 
vectors $\{u_j\}$. 
The matrix $U$ built by $\{u_j\}$ as column vectors is a
truncated unitary with its Moore-Penrose pseudoinverse~\cite{Moore1920}
defined by $U^{-1}=U^\dagger$.
Here, however, the presence of the singular values and their reciprocals
in Eqs.~\eqref{eq:T_for} and \eqref{eq:T_back} leads to transformation matrices
and pseudoinverses which are not related by Hermitian conjugation
$\mathcal{T}^{-1}\neq\mathcal{T}^\dagger$. 
In fact, $\mathcal{T}^{-1}$ is not the unique Moore-Penrose pseudoinverse
of $\mathcal{T}$, i.e., the column vectors of $\mathcal{T}^{-1}$ do
not span the same space as the conjugates of the row vectors of $\mathcal{T}$.

For the purpose of interpretation, this implies that the relevant subspace 
to which inner bonds of PTs correspond is slightly ambiguous, as it can be
identified with the column space of either $\mathcal{T}^{-1}$ or 
$\mathcal{T}^\dagger$.
For practical applications, such as the extraction of information from the
inner bonds, it entails that the pseudoinverse $\mathcal{T}^{-1}$ 
contains additional information and cannot be reconstructed from
$\mathcal{T}$ alone.

Moreover, we find from Eq.~\eqref{eq:Ttotal} that
%, after summing over $j_l^{(k)}$, 
the transformation matrices 
$\mathcal{T}$ and $\mathcal{T}^{-1}$ themselves have the structure of 
half-open MPOs, where each environment mode $k$ corresponds to a site 
with outer bond $\beta^{(k)}_l$, while $d^{(k)}_l$ describe the inner bonds.
The last inner bond $d^{(N_E)}_l$ of the tranformation matrix MPO 
remains dangling and is identified the inner bond $d_l$ of the final PT-MPO
at step $l$.
The dangling bond can be closed by multiplying with $\rho^{\alpha_l}_{d_l}$,
the extended density matrix obtained at step $l$ during the ACE
algorithm by interation~\eqref{eq:iter}. The result
\begin{align}
&\rho_{\alpha_l,\beta_l^{(1)},\dots,\beta_l^{(N_E)}} = \sum_{d_l^{(N_E)}}
\mathcal{T}^{-1}_{(\beta_l^{(1)},\dots,\beta_l^{(N_E)}, d_l^{(N_E)}} 
\rho^{\alpha_l}_{d_l^{(N_E)}}
\nonumber\\
&= \sum_{d_l^{(1)},\dots,d_l^{(N_E)}}
\bigg(\prod_{k=1}^{N_E}
\overleftrightarrow{T}^{-1}_{(d_{l}^{(k-1)},\beta^{(k)}_l),d_l^{(k)}}\bigg)
\rho^{\alpha_l}_{d^{(N_E)}_l}
\label{eq:rhoMPO}
\end{align}
is an MPO representation of the total system and environment density matrix.

\subsection{\label{sec:closures}Extraction of environment observables}
As outlined in Eq.~\eqref{eq:Oextract_formal} and in line with 
Eq.~\eqref{eq:rhoMPO}, the knowledge of $\mathcal{T}^{-1}$ enables the
extraction of environment observables as well as mixed system-environment
observables.
Starting from Eq.~\eqref{eq:Osupav}, the expectation value of a
general mixed system environment observable $\hat{O}$ with Liouville space 
representation $o_{\alpha,\beta}$ given by Eq.~\eqref{eq:Osupdef}
is obtained by
\begin{align}
\langle\hat{O}(t_l)\rangle =& 
\sum_{\alpha_l,\beta_l}  o_{\alpha_l,\beta_l}\rho_{\alpha_l,\beta_l}
\approx
\sum_{\substack{\alpha_l,\beta_l\\d_l,\beta'_l}} 
o_{\alpha_l,\beta_l} \mathcal{T}^{-1}_{\beta_l, d_l} 
\mathcal{T}_{d_l,\beta'_l}\rho_{\alpha_l,\beta'_l}
\nonumber\\
=& \sum_{\alpha_l, d_l} \mathfrak{o}^{\alpha_l}_{d_l} \rho^{\alpha_l}_{d_l}
\label{eq:oad_av}
\end{align}
with observable closure
\begin{align}
\label{eq:oad_def}
\mathfrak{o}^{\alpha_l}_{d_l}=\sum_{\beta_l} 
o_{\alpha_l,\beta_l} \mathcal{T}^{-1}_{\beta_l, d_l}.
\end{align}

Here, two issues arise if Eqs.~\eqref{eq:oad_av} and \eqref{eq:oad_def} are
to be used to extract environment observables in practice:
First, the matrices $\mathcal{T}^{-1}_{\beta_l,d_l}$ change from time 
step to time step. Even though they can be represented in MPO form by
Eq.~\eqref{eq:Ttotal_b}, storing them for all time steps $l$ requires 
large amounts of memory.
This can be circumvented by fixing an operator $\hat{O}$ at the start
and updating the corresponding $\mathfrak{o}^{\alpha_l}_{d_l}$ in every expansion and
compression step of the ACE algorithm in the same way that 
$\mathcal{T}^{-1}_{\beta_l,d_l}$ would be updated, 
namely by multiplying with 
$\delta_{(d_l,\beta_l^{(K)}),\tilde{d}_l}$ as in Eq.~\eqref{eq:Texpand_b},
$\overleftarrow{T}^{-1}_{d_l,j_l}$ as in Eq.~\eqref{eq:Ttrafo_for_b},
and $\overrightarrow{T}^{-1}_{d_l,j_l}$ as in Eq.~\eqref{eq:Ttrafo_back_b}.

The second issue is that $o_{\alpha_l,\beta_l}$ itself can be unwieldy 
in general, as $\beta_l$ runs over a complete basis set for the full 
environment Liouville space. 
Here, we therefore limit the discussion to operators of the form
\begin{align}
\hat{O} = \sum_{k=1}^{N_E} \hat{A}\otimes \hat{O}^{(k)},
\label{eq:O_decompose}
\end{align}
where $\hat{A}$ acts on the system Hilbert space $\mathcal{H}_S$ and
$\hat{O}^{(k)}$ acts on the $k$-th environment mode. 
We define the corresponding Liouville operators as
\begin{align}
o_{\alpha,\beta^{(1)},\dots,\beta^{(N_E)}}=
o^\alpha \sum_{k=1}^{N_E} o^{(k)}_{\beta^{(k)}}.
\label{eq:Osum}
\end{align}
Then, the corresponding observable closure $\mathfrak{o}^{\alpha_l}_{d_l}$ 
defined in Eq.~\eqref{eq:oad_def} is obtained alongside 
the PT in the ACE algorithm by the iteration
\begin{align}
\mathfrak{o}_{d_l^{(k)}}=&
\sum_{d_l^{(k-1)}}\sum_{\beta_l^{(k)}}\Big( 
\mathfrak{o}_{d_l^{(k-1)}} \mathfrak{I}_{\beta_l^{(k)}}
+ q_{d_l^{(k-1)}} o^{(k)}_{\beta^{(k)}}\Big)
\nonumber\\&\times
\overleftrightarrow{T}^{-1}_{(d_{l}^{(k-1)},\beta_l^{(k)}),d_l^{(k)}},
\label{eq:o_iter}
\end{align}
where the first summand describes the expansion and transformation of
$\mathfrak{o}_{d_l^{(k-1)}}$, 
which accounts for contributions to the overall
observable from modes $1,2,\dots,k-1$,
while the second summand adds the contribution $o^{(k)}_{\beta^{(k)}}$ from environment mode $k$ 
corresponding to the $k$-th term of the sum in Eq.~\eqref{eq:Osum}.
The symbol $\mathfrak{I}_{\beta_l^{(k)}}$ accounts for the trace over 
the $k$-th environment mode as introduced below Eq.~\eqref{eq:OS_Ifrak},
and
$q_{d_l^{(k)}}$ are the closures at intermediate time steps corresponding
to a trace over environment modes $1,2,\dots k$, 
which are also obtained iteratively by
\begin{align}
q_{d_l^{(k)}}=\sum_{d_l^{(k-1)}}\sum_{\beta_l^{(k)}}  q_{d_l^{(k-1)}}
\mathfrak{I}_{\beta_l^{(k)}}
\overleftrightarrow{T}^{-1}_{(d_{l}^{(k-1)},\beta_l^{(k)}),d_l^{(k)}}.
\label{eq:q_iter}
\end{align}
The final observable closure $\mathfrak{o}^{\alpha_l}_{d_l}$ is identified with
the result of the iteration~\eqref{eq:o_iter} multiplied with the system part of the operator
$\mathfrak{o}^{\alpha_l}_{d_l}=o^{\alpha_l} \mathfrak{o}_{d_l^{(N_E)}}$.
Moreover, identifying $q_{d_l}=q_{d_l^{(N_E)}}$ provides an alternative way
to that described in Ref.~\cite{ACE} 
for obtaining trace closures $q_{d_l}$ used to extract the reduced system
density matrix from PT-MPO calculation at intermediate time steps via 
$\bar{\rho}_{\alpha_l}=\sum_{d_l}q_{d_l}\rho^{\alpha_l}_{d_l}$.

\section{\label{sec:examples}Examples}
\subsection{Proof of principle: Environment spins}
First, we demonstrate a minimal example for the extraction of
environment observables on the basis of a central spin model.
We consider a central spin $\mathbf{S}$
coupled to a bath of $N_E$ spins $\mathbf{s}^{(k)}$ 
via the Heisenberg interaction
\begin{align}
H_E=&\sum_{k=1}^{N_E} J\, \mathbf{S}\cdot \mathbf{s}^{(k)}.
\end{align}
Here, we assume that the central spin is initially polarized along the $x$ 
direction $S_x(t=0)=\frac 12$, $S_y(t=0)=S_z(t=0)=0$, while 
the $N_E=25$ environment spins are initially randomly oriented. 
The central spin $\mathbf{S}$ is obtained as usual with ACE
from the reduced system density matrix, while the sum of environment spins
$\sum_k \mathbf{s}^{(k)}$ is obtained as an environment observable 
by tracking the corresponding observable closure during MPO compression.
The PT-MPO is calculated for convergence
parameters $(J/\hbar)\Delta t=0.01$ and $\epsilon=10^{-11}$, for which
the full environment Liouville space dimension $4^{25}\approx 10^{15}$ is 
compressed to a maximal bond dimension of $136$.

\begin{figure}
\includegraphics[width=\linewidth]{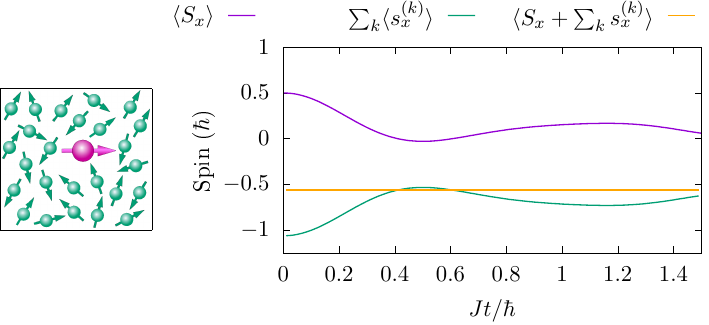}
\caption{Dephasing of a central spin $S_x$, initially polarized along the $x$ axis,
in an ensemble of $N_E=25$ environment spins, initially randomly oriented 
(see sketch on the left). The overall environment spin projection $\sum_k s^{(k)}_x$ is
extracted from the inner degrees of freedom of the PT-MPO. The conservation
of the total spin $S_x+\sum_k s^{(k)}_x$ is correctly reproduced.
\label{fig:spins}}
\end{figure}

The ensuing spin dynamics ($x$-component) is depicted in Fig.~\ref{fig:spins}. 
The central spin $S_x$ dephases in the bath of environment spins, 
yet the dephasing is incomplete due to the finite size of the spin bath.
The dynamics of the collective environment spin $\sum_k s^{(k)}_x$ changes accordingly
as the total central plus environment spin $S_x+\sum_k s^{(k)}_x$
is conserved by the Hamiltonian.
This is correctly reproduced in Fig.~\ref{fig:spins} by our numerical approach
demonstrating the successful extraction of environment observables.

\subsection{Mixed system-environment observables: Currents}
\begin{figure}
\includegraphics[width=0.99\linewidth]{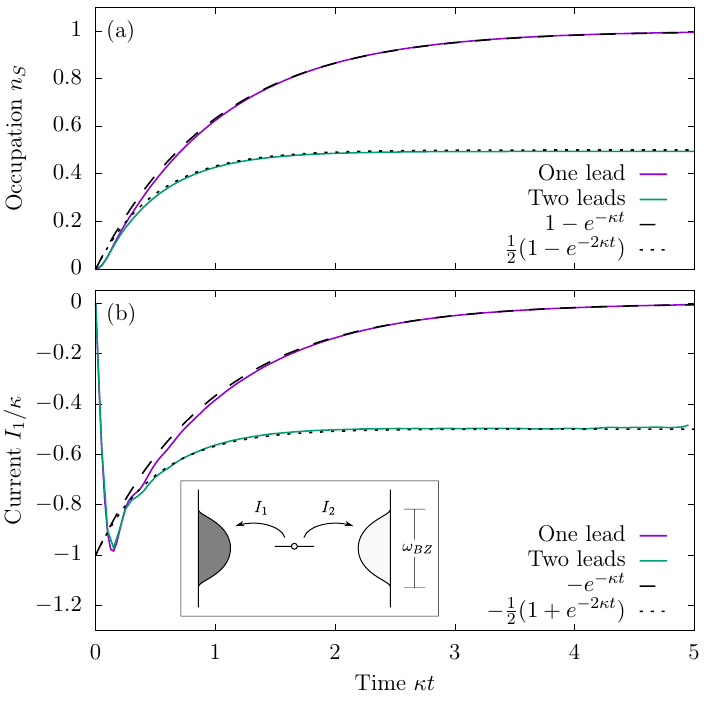}
\caption{\label{fig:current}%
Occupations (a) and currents (b) in charge transport 
between a single site and one or two metallic leads at 
chemical potentials $\mu_{1/2}=\pm \infty$, as depicted in the inset of (b). 
Dashed and dotted black lines depict results in the Markov limit for the single lead case and the two lead case, respectively.
}
\end{figure}

To infer physical insights from measurement of currents 
through zero-dimensional quantum structures like single molecules 
or quantum dots, a comparison with theoretical predictions is 
highly desirable~\cite{beyondMarcus}.
A typical scenario for charge transport involves a single quantum site,
which is described by a two-level system (the site being either occupied or not),
and which is coupled to two metallic leads at different chemical potentials
$\mu_1>\mu_2$. 
The leads are modelled as fermionic baths.
The total Hamiltonian is
$H=H_S + H_{E_1}+H_{E_2}$ with $H_{E_i}=\sum_k H_{E_i}^{(k)}$ and
\begin{align}
\label{eq:Hfermion}
H_{E_i}^{(k)}=&  \hbar\omega_{i,k} c^\dagger_{i,k}c_{i,k} +
 \hbar g_{i,k} \big(c^\dagger_{i,k} c_S + c^\dagger_{S} c_{i,k} \big),
\end{align}
where $c^{(\dagger)}_{i,k}$ denote fermionic annihilation (creation) operators for the $k$-th mode of the $i$-th environment,
$\hbar\omega_{i,k}$ and $g_{i,k}$ are the corresponding energies and 
coupling constants, and
$c^\dagger_S$ and $c_S$ describe fermionic operators of the system mode.
We set $H_S=0$, and assume that the
central site is initially unoccupied, the first lead is 
completely filled and the second lead is completely empty 
(chemical potentials $\mu_{1/2}=\pm \infty$). 
The energies and couplings for both baths are obtained by uniformly 
discretizing the spectral densities
$J_i(\omega)=\sum_{k}g_{i,k}^2\delta(\omega-\omega_{i,k})$
of the shape of a bump function
\begin{align}
J_i(\omega)=&\begin{cases}\frac{\kappa}{2\pi}
\exp\Big(1-\frac{1}{1-(2\omega/\omega_{BW})^2}\Big), 
& |\omega|<\frac 12\omega_{BW}, \\ 
0,& \textrm{else}.
\end{cases}
\end{align}
This is a smooth function with compact support on an interval with bandwidth
$\omega_{BW}$ (see inset in Fig.~\ref{fig:current}).

The particle current flowing from the central site into lead $i$ is equal to 
the change of the total occupations of the lead due to coupling to the site.
From the Heisenberg equations of motion we find
\begin{align}
I_i:=&\frac{\partial}{\partial t} \sum_k\langle c^\dagger_{i,k}c_{i,k}\rangle
=\frac{i}{\hbar}\sum_k \langle [H,  c^\dagger_{i,k}c_{i,k}] \rangle
\nonumber\\=&
2\sum_k g_{i,k} \textrm{Im}\big\{\langle c^\dagger_{i,k}c_S \rangle\big\}.
\label{eq:expression_current}
\end{align}
This is a mixed system-environment observable, which depends on correlations between the system and the environment. At first glance, this observable as well as the environment Hamiltonian in Eq.~\eqref{eq:Hfermion} seem decomposable into sums of terms each involving only creation and annihilation operators of a single environment mode, as we require for the  environment observable extraction described in this article. Note, however, that fermionic operators obey the canonical anticommutator relations, whereas the decomposition into independent modes so far implicitly assumed that operators for different modes commute. 
As described in more detail in Ref.~\cite{combine_tree}, PT-MPOs obeying the proper fermionic anticommutator relations can be obtained by a slight modification of ACE based on a Jordan-Wigner transformation. Here, we briefly summarize the main aspects while using a formulation that facilitates the extraction of the current $I_1(t)$ in a setup with up to two environments.

First, the Jordan-Wigner transformations requires an ordering of fermionic modes. Here, we assume the order
\begin{align}
\label{eq:JWorder}
(1,1),(1,2),\dots,(1,N_{E_1}), (S), (2,N_{E_2}), \dots, (2,2),(2,1),
\end{align}
where (S) denotes the system mode and $(i,k)$ denotes the $k$-th mode of the $i$-th environment. The anticommutation relations remain fulfilled when the fermionic operators are replace by spin-1/2 climbing operators 
\begin{subequations}
\begin{align}
c_{i,k}=& P^{[(1,1),(i,k-1)]}\sigma^-_{i,k}, \\
c^\dagger_{i,k}=& P^{[(1,1),(i,k-1)]}\sigma^+_{i,k},
\end{align}
\end{subequations}
where
\begin{align}
P^{[(i_1,k_1),(i_2,k_2)]} = & \prod_{(i,k)=(i_1,k_1)}^{(i_2,k_2)} (-\sigma^z_{i,k})
\end{align}
is the product of parity operators from mode $(i_1,k_1)$ to $(i_2,k_2)$ in the order given by Eq.~\eqref{eq:JWorder}, which takes the value +1 (-1) if the modes in the range have an even (odd) number of occupations.
Applying this transformation to the Hamiltonian 
in Eq.~\eqref{eq:Hfermion}, one can show that
\begin{align}
\label{eq:H_to_Htilde}
H_{E_i}^{(k)} &=  \big(P^{[(1,1),(i,k-1)]}_1\big)^{\sigma^+_S\sigma^-_S} 
\tilde{H}_{E_i}^{(k)} \big(P^{[(1,1),(i,k-1)]}_1\big)^{\sigma^+_S\sigma^-_S},
\end{align}
where 
\begin{align}
\label{eq:Htilde}
\tilde{H}_{E_i}^{(k)}=&  \hbar\omega_{i,k} \sigma^+_{i,k}\sigma^-_{i,k} +
 \hbar g_{i,k} \big(\sigma^+_{i,k} \sigma^-_S + 
 \sigma^+_{S} \sigma^-_{i,k} \big)
\end{align}
is the spin analogue of the fermionic Hamiltonian $H_{E_i}^{(k)}$.

Now, while the fermionic mode Hamiltonians individually contain non-local terms via the parity operators, it can be shown that these non-local terms cancel when the propagators of all modes are combined, leaving only local parity terms~\cite{combine_tree}. Specifically, representing the PT-MPO matrices for environment $i$ using lossy compression matrices as well as the symmetric Trotter decomposition of the full environment propagator in Eq.~\eqref{eq:symTrotterK}, one finds
\begin{align}
&\mathcal{Q}^{[i]}
\nonumber\\&= \mathcal{T} e^{\mathcal{L}_{E_i}^{(N_{E_i})}\frac{\Delta t}2} \dots  
e^{\mathcal{L}_{E_i}^{(2)}\frac{\Delta t}2}e^{\mathcal{L}_{E_i}^{(1)}\Delta t}
e^{\mathcal{L}_{E_i}^{(2)}\frac{\Delta t}2} \dots
 e^{\mathcal{L}_{E_i}^{(N_{E_i})}\frac{\Delta t}2}\mathcal{T}^{-1}
\nonumber\\&
=\mathcal{T} \mathcal{B}^{[i,N_{E_i}]} \dots \mathcal{B}^{[i,2]} \mathcal{B}^{[i,1]} \mathcal{B}^{[i,1]} \mathcal{B}^{[i,2]}\dots \mathcal{B}^{[i,N_{E_i}]}
\mathcal{T}^{-1},
\label{eq:Trotter_fermion}
\end{align}
%where $\mathcal{L}_{E_i}^{(k)}$ and  $\tilde{\mathcal{L}}_{E_i}^{(k)}$ are the Liouvillians corresponding to fermionic and spin Hamiltonians $H_{E_i}^{(k)}$ and $\tilde{H}_{E_i}^{(k)}$, respectively. 
%Thus, Eq.~\eqref{eq:Trotter_fermion} mostly contains independent blocks
where
\begin{align}
\mathcal{B}^{[i,k]}:=& e^{\tilde{\mathcal{L}}_{E_i}^{(k)} \frac{\Delta t}2} 
\Big[\big(-\sigma^z_{i,k}\big)^{\sigma^+_S\sigma^-_S}\otimes
\big(-\sigma^z_{i,k}\big)^{\sigma^+_S\sigma^-_S}\Big]
\label{eq:prop_fermion}
\end{align}
are the local environment mode propagators corresponding to the spin Hamiltonians $\tilde{H}_{E_i}^{(k)}$ in Eq.~\eqref{eq:Htilde} modified by local parity operators $\big(-\sigma^z_{i,k}\big)^{\sigma^+_S\sigma^-_S}$. 
To summarize, PT-MPOs correctly accounting for fermionic anticommutation can be obtained simply by replacing the individual mode propagators $\mathcal{B}$ in Eq.~\eqref{eq:BQB} by corresponding $\mathcal{B}^{[i,k]}$ in Eq.~\eqref{eq:prop_fermion}.

To extract the particle current via the inner bonds of the fermionic PT-MPO, we apply the Jordan-Wigner transform to Eq.~\eqref{eq:expression_current}, which yields
\begin{align}
I_i:=2\sum_k g_{1,k}\textrm{Im}\big\{\langle\sigma^+_{i,k} P^{[k+1,N_{E_i}]}_1 \sigma^-_S \rangle \big\}.
\end{align}
Again the non-local parity terms make adaptations necessary. This is achieved by replacing the iteration in Eq.~\eqref{eq:o_iter} by
\begin{align}
\mathfrak{o}_{d_l^{(k)}}=&
\sum_{d_l^{(k-1)}}\sum_{\beta_l^{(k)},\xi,\eta} \delta_{\beta_l^{(k)},(\xi,\eta)}
\Big( \mathfrak{o}_{d_l^{(k-1)}}\langle \eta|-\sigma^z|\xi\rangle
\nonumber\\&+ q_{d_l^{(k-1)}} \langle \eta| \sigma^+ |\xi\rangle \Big)
\overleftrightarrow{T}^{-1}_{(d_{l}^{(k-1)},\beta_l^{(k)}),d_l^{(k)}}.
\end{align}
%\begin{align}
%\mathfrak{o}_{d_l^{(k)}}=&
%\sum_{d_l^{(k-1)}}\sum_{\beta_l^{(k)}}\Big( 
%\mathfrak{o}_{d_l^{(k-1)}} \mathfrak{I}_{\beta_l^{(k)}}
%+ p_{d_l^{(k-1)}} o^{(k)}_{\beta^{(k)}}\Big)
%\nonumber\\&\times
%\overleftrightarrow{T}^{-1}_{(d_{l}^{(k-1)},\beta_l^{(k)}),d_l^{(k)}},
%\end{align}
%where the identity closure $q_{d_l^{(k)}}$ is replaced by the parity closure $p_{d_l^{(k)}}$, which itself can be constructed in parallel with $\mathfrak{o}_{d_l^{(k)}}$ by an iteration similar to Eq.~\eqref{eq:q_iter}
%\begin{align}
%p_{d_l^{(k)}}=&\sum_{d_l^{(k-1)}}\sum_{\beta_l^{(k)}}  p_{d_l^{(k-1)}}
%\delta_{\beta_l^{(k)},(\xi,\eta)} \langle \eta|-\sigma^z|\xi\rangle
%\nonumber\\&\times
%\overleftrightarrow{T}^{-1}_{(d_{l}^{(k-1)},\beta_l^{(k)}),d_l^{(k)}}.
%\label{eq:p_iter}
%\end{align}
With this observable closure, we are now in the position to extract the fermionic particle current $I_i(t)$ as a mixed system-environment observable from the inner bonds of the fermionic PT-MPO.

Finally, we point out that the numerically exact simulation in the presence of two generally non-Markovian environments is possible by propagating the time evolution with two PT-MPOs, which are calculated independently of each other. This was utilized already for bosonic environments, e.g., in Refs.~\cite{CoopWiercinski, twobath}. Here, the ordering used for the Jordan-Wigner transformation in Eq.~\eqref{eq:JWorder} ensures that no additional non-local parity operators coupling the two environments emerge.
The fact that this treatment remains numerically exact 
can be seen straightforwardly in the notation developed here: 
Using transformation matrices $\mathcal{T}^{[i]}$ and PT-MPO matrices 
$\mathcal{Q}^{[i]}=\mathcal{T}^{[i]} e^{\mathcal{L}_{E_i}\Delta t} 
\big(\mathcal{T}^{[i]}\big)^{-1}$ for bath $i$, we find that
\begin{align}
\label{eq:Qtwobath}
\mathcal{Q}^{[1]}\mathcal{Q}^{[2]}=&
\mathcal{T}^{[1]}\mathcal{T}^{[2]} e^{\mathcal{L}_{E_1}\Delta t}
e^{\mathcal{L}_{E_2}\Delta t} \big(\mathcal{T}^{[2]}\big)^{-1} 
\big(\mathcal{T}^{[1]}\big)^{-1}
\nonumber\\ = & 
\mathcal{T} e^{(\mathcal{L}_{E_1}+\mathcal{L}_{E_2})\Delta t}
\mathcal{T}^{-1} +\mathcal{O}(\Delta t^2)
\end{align}
describes the joint influence of both environments up to a controlled 
Trotter error. Note that the order of the Trotter error can be reduced when alternating
the order of multiplication (with respect to the system indices) 
for subsequent time steps, i.e. setting 
$\mathcal{Q}=\mathcal{Q}^{[1]}\mathcal{Q}^{[2]}$ for odd and 
$\mathcal{Q}=\mathcal{Q}^{[2]}\mathcal{Q}^{[1]}$ for even times steps,
because this amounts to a symmetric Trotter decomposition of a propagation
over two time steps. For a consistent accuracy, observables at odd times
steps are disregarded.

In Fig.~\ref{fig:current}(a) and (b) we show the system site occupations 
and the particle current from the system to the first lead, respectively, 
for simulations involving either only the first lead or both leads. 
The PT-MPOs are calculated using parameters
$\omega_{BW}=64\kappa$, $N=128$, $\Delta t=0.025/\kappa$, and 
$\epsilon=10^{-7}$.

In the case of a single lead with large chemical potential, carriers flow
from lead to the system site (negative $I_1$) until the latter is 
fully populated. The dynamics is largely in agreement with the 
analytical behavior $n_S(t)=1-e^{-\kappa t}$ 
in the Markov limit, which is expected to hold as
the bath correlation time $\sim 1/\omega_{BW}$ is short compared to the
relaxation time $1/\kappa$. 
Due to the conservation of particle number and the coupling to only
a single lead, the current can be obtained from the
system evolution via
$I_1(t)=-\frac{\partial}{\partial}n_S(t)$, which is 
$I_1(t)=-\kappa e^{-\kappa t}$ in the Markov limit.

For a site coupled to two leads, changes of site occupations are due to
currents to either lead, which generally obfuscates the relation between currents
through one of the leads and system observables. 
This is where extracting the current via the mixed system-environment
observable defined in Eq.~\eqref{eq:expression_current} is useful.
In the present scenario where initially one lead is fully occupied and one lead is completely empty, the dynamics can be compared to its Markovian limit, where it is governed by the rate equation
$\frac{\partial}{\partial t} n_S=-I_1(t)-I_2(t)$, where
$I_1(t)=\kappa(n_S(t)-1)$ and $I_2(t)=\kappa n_S(t)$ are the Markovian currents to environment 1 and 2, respectively. This equation of motion is solved by system occupations $n_S(t)=\frac 12(1-e^{-2\kappa t})$, from which we get the current
$I_1(t)=-\frac{\kappa}2(1+e^{-2\kappa t})$.
As can be seen in Fig.~\ref{fig:current}, for the case of two leads, simulated site occupations as well as currents also match the Markovian prediction well after a short initial phase on the timescale $\sim 1/\omega_{BW}$. 

Summarizing, we have demonstrated that inner bonds of PT-MPOs can be used 
to extract mixed system-environment operators, which facilitate, 
e.g., the analysis of currents. 
In principle, the combination of multiple PT-MPOs 
also makes it possible to address more complex questions, such as the 
impact of strong phonon coupling on charge and excitation transport. 
As this topic requires a more detailed analysis, we leave it for future work.

\subsection{Convergence of different observables: 
Application in photon emission}
Because the transformation matrices $\mathcal{T}$ are lossy, 
i.e. rank-reducing, there is no \emph{a priori} guarantee that a given
environment observable can be extracted faithfully by the
corresponding pseudoinverse $\mathcal{T}^{-1}$. Moreover,
the accuracies may be different for different environment observables.
This fact can be utilized to probe the information content of inner bonds
of PT-MPOs by numerically testing the convergence of different environment 
observables as a function of the parameters controlling MPO compression. 
The insights gained by this process then helps to identify alternative ways to
extract environment observables with a higher degree of accuracy.

\begin{figure}
\includegraphics[width=0.99\linewidth]{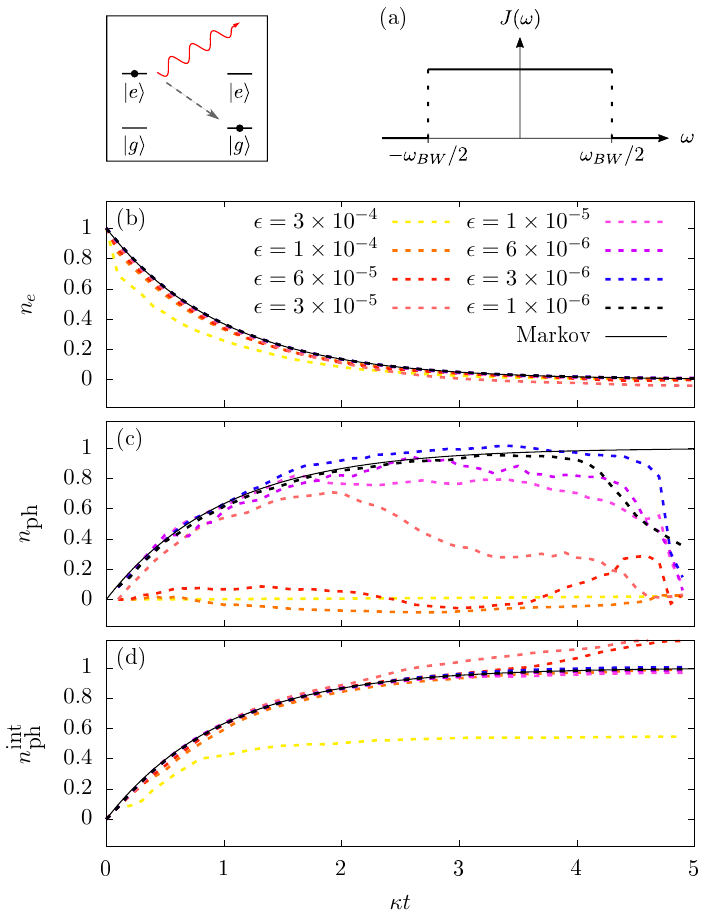}
\caption{Photon emission from an initially excited two-level system (see sketch). 
This is described microscopically by a multi-mode Jaynes-Cummings model with flat spectral density over a bandwidth $\omega_{BW}$ as depicted in (a).
Two-level system occupation $n_e$ (b) and 
emitted photon number $n_\textrm{ph}$, where the
photon number is either extracted directly (c) or by integrating 
its equation of motion, whose driving term is obtained by extracting
system-environment correlations (d) as in Eq.~\eqref{eq:n_ph_int}.
\label{fig:radiative}}
\end{figure}

We explore this on the example of radiative decay from a two-level 
quantum emitter. The light-matter interaction is given by
\begin{align}
H_E=& \sum_k \hbar \omega_k a^\dagger_k a_k + 
\sum_k \hbar g_k \big( a^\dagger_k |g\rangle\langle e|
+ a_k|e\rangle\langle g| \big),
\end{align}
where $a^\dagger_k$ and $a_k$ are photon creation and annihilation operators 
and $|g\rangle$ as well as $|e\rangle$ are ground and excited states of
the emitter, respectively.
Photon mode energies $\hbar\omega_k$ are chosen to uniformly discretize 
an interval $[-\hbar\omega_{BW}/2, \hbar\omega_{BW}/2]$ 
with frequency bandwidth $\omega_{BW}$ using $N_E$ modes.
The couplings are obtained by $g_k = \sqrt{J(\omega_k) \omega_{BW}/N_E}$
with a flat spectral density $J(\omega)=\kappa/(2\pi)$ [see Fig.~\ref{fig:radiative}(a)].
For a large enough bandwidth $\omega_{BW}\gg \kappa$, 
the photon environment is Markovian~\cite{ACE}
and describes radiative decay of the emitter excitation 
$n_e=\langle \big( |e\rangle\langle e|\big)\rangle$ 
with rate $\kappa$. Thus, for an initially excited emitter $n_e(0)=1$ the
excitation decays as
$n_e^{\textrm{Markov}}\approx e^{-\kappa t}$.
As the Hamiltonian conserves the total number of excitations, it follows that
the number of photons emitted into the environment 
$n_\textrm{ph}=\sum_k \langle a^\dagger_k a_k\rangle$ is
$n_\textrm{ph}^{\textrm{Markov}}= 1-e^{-\kappa t}$.

We use this test case to assess the convergence of the two-level system
excitations $n_e$ as well as the environment observable $n_\textrm{ph}$.
The numerically calculated dynamics is depicted in 
Fig.~\ref{fig:radiative}(b) and (c) 
for different MPO compression thresholds $\epsilon$,
fixed bandwidth $\omega_{BW}=200\kappa$, number of modes $N_E=400$,
and time discretization $\Delta t=0.05/\kappa$.

The system observable $n_e$ converges quickly and becomes 
virtually indistinguishable from the Markovian result for thresholds 
$\epsilon=10^{-5}$ and smaller. 
In contrast, the extraction of the photon number $n_\textrm{ph}$ as 
an environment observable is found to be more unstable and converges 
much more slowly. 
This is because the compression of the PT-MPO in ACE
is only designed to accurately reproduce the reduced system density matrix 
and, hence, system observables. 
If the compression is optimal in this sense, the accuracy of the extracted 
environment observables should be determined by how important they are
for influencing system observables in future time steps.

To test this hypothesis, we consider the Heisenberg equations of motion 
for emitter occupations
\begin{align}
\frac{\partial}{\partial t} n_e=&
\frac i\hbar \langle \big[ H_S+H_E , |e\rangle\langle e|\big] \rangle 
\nonumber\\
=&\frac i\hbar \langle \big[ H_S, |e\rangle\langle e|\big] \rangle
+ 2\sum_k g_k \textrm{Im}\Big\{\langle \big(|e\rangle\langle g| a_k\big) \rangle\Big\}.
\label{eq:eom_ne}
\end{align}
Note that the system evolution is directly driven by system-environment 
correlations $\langle \big(|e\rangle\langle g| a_k\big) \rangle$ but not
by the photon number $n_\textrm{ph}$ itself. The latter affects 
the system only indirectly by influencing the evolution of 
the system-environment correlations.
Because a finite time is needed for the influence of $n_\textrm{ph}$ on 
the correlations to result in a measurable effect on the system, 
$n_\textrm{ph}$ is reproduced most accurately in
Fig.~\ref{fig:radiative}(c) at earlier points in time. In contrast, 
$n_\textrm{ph}$ at later time steps can no longer affect the state of the 
system within the remaining propagated time in the simulation, so the
extracted value of $n_\textrm{ph}$ at the last few time steps remains
unreliable even for small thresholds $\epsilon=10^{-6}$.

This explanation also provides a hint on how to construct a more accurate 
scheme to extract the total emitted photon number: Considering the 
equation of motion
\begin{align}
\frac{\partial}{\partial t} n_\textrm{ph}=&
\frac i\hbar \langle \big[ H_E , \sum_k a^\dagger_k a_k  \big] \rangle 
=-2\sum_k g_k \textrm{Im}\Big\{\langle \big(|e\rangle\langle g| a_k\big) \rangle\Big\},
\label{eq:eom_nph}
\end{align}
we find that the photon number is driven by the same system-environment
correlations as the emitter occupation. As these correlations influence
the system directly, they are likely better reproduced when 
extracted from the inner bonds of the PT-MPO than $n_\textrm{ph}$ itself.
This suggests obtaining the photon number by integrating Eq.~\eqref{eq:eom_nph}
\begin{align}
\label{eq:n_ph_int}
n_\textrm{ph}^\textrm{int}(t)=&
\int\limits_0^t d\tau\Big(\frac{\partial}{\partial \tau} n_\textrm{ph}(\tau)\Big)=
\int\limits_0^t d\tau\, 
\sum_k g_k \langle \big(|e\rangle\langle g| a_k \big) \rangle_\tau,
\end{align}
where $\langle \big(|e\rangle\langle g| a_k \big) \rangle_\tau$ are the
system-environment correlations at time $\tau$, which are extracted
via the corresponing observable closures.

The result of this approach is depicted in Fig.~\ref{fig:radiative}(d). 
Indeed, the convergence with respect to the compression threshold $\epsilon$
is much faster compared to the direct extraction of $n_\textrm{ph}$ 
in Fig.~\ref{fig:radiative}(c). This corroborates the argument that
the inner bonds of PT-MPOs convey more information about first-order 
system-environment correlations that directly affect the system 
dynamics compared to environment observables with a more indirect influence
on the system. The hierarchy of Heisenberg equations of motion starting 
from system observables provide a useful basis for a qualitative estimation
of the expected accuracy of environment observable extraction via inner bonds 
of PT-MPOs.

%\subsection{Driven dynamics in structured environment}
\begin{figure}
\includegraphics[width=\linewidth]{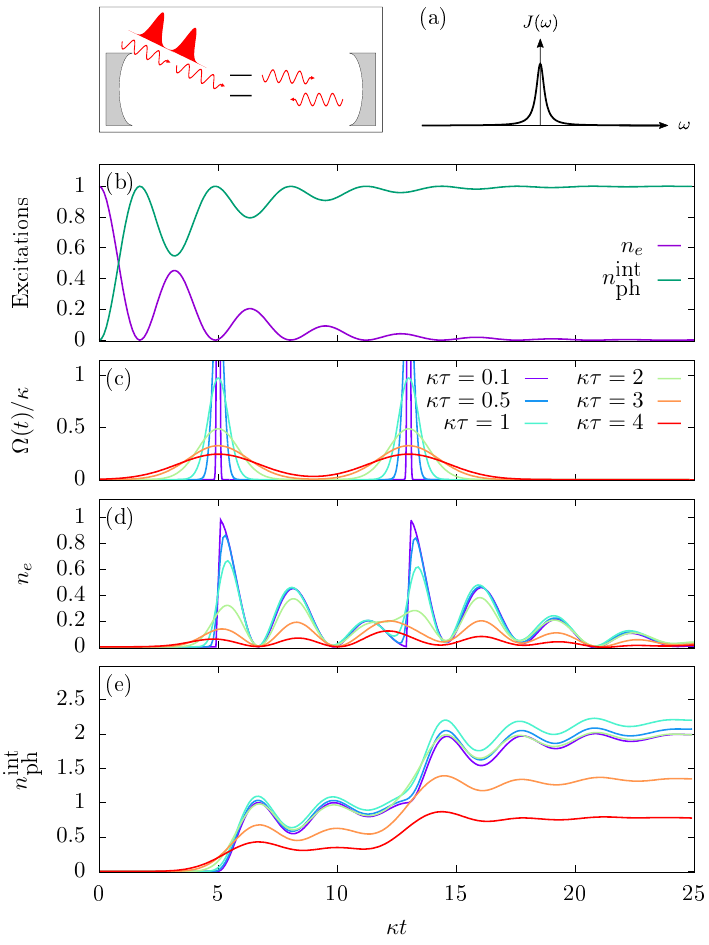}
\caption{Radiative decay from a two-level quantum emitter
into a structured photon environment such as a lossy single-mode microcavity (see sketch), which is modeled with a Lorentzian spectral density $J(\omega)$ (a). 
(b) Free radiative decay of initially prepared emitter populations $n_e$.
The photon number $n_\textrm{ph}^\textrm{int}$ is extracted by integrating
system-environment correlations as in Eq.~\eqref{eq:n_ph_int}.
(d) Emitter populations and (e) emitted photon number for the same system
driven by two Gaussian pulses depicted in (c). Colors correspond
to different pulse widths $\tau_1=\tau_2=\tau$.
\label{fig:lorentzian}}
\end{figure}

The nearly Markovian environment with a flat spectral density provided 
an ideal test case for different observable  extraction schemes due
to the availability of analytical solutions.
Numerically exact open quantum systems approaches are most
useful in cases of non-Markovian environments, where no analytical solutions
exist.
If these systems are also externally driven, inferring environment observables
from conservation laws is often no longer possible.
We now show how the approach laid out above can be applied to investigate
photon emission in a model of 
a two-level emitter coupled to a structured, non-Markovian photonic environment 
with spectral density given by a Lorentzian [see Fig.~\ref{fig:lorentzian}(a)]
\begin{align}
\label{eq:J_lorentzian}
J(\omega)=&\frac{\kappa^2}{\pi}\frac{\gamma}{\omega^2 + \gamma^2},
\end{align}
where $\kappa$ determines the overall interaction strength and $\gamma$ 
denotes the width of the Lorentzian. 
The corresponding PT-MPO is calculated restricting the spectral density 
to a frequency interval $\omega_{BW}=200\kappa$, which is discretized by
$N_E=2000$ photon mode with a Hilbert space containing up to 2 photons per mode.
Furthermore, we use parameters $\gamma=\frac 14\kappa$, $\Delta t=0.05/\kappa$,
and $\epsilon=10^{-7}$. %, and propagate over a total time $t_e=25/\kappa$.

In Fig.~\ref{fig:lorentzian}(b), we show the free ($H_S=0$) 
emission dynamics of an initially prepared emitter excitation.
The strongly peaked structure in the environment results in underdamped
oscillations of emitter population $n_e$. The emitted photon number
$n_\textrm{ph}^\textrm{int}$ extracted by integrating the system-emitter
correlations as in Eq.~\eqref{eq:n_ph_int} mirrors the behavior of the
emitter population, 
since the light-matter coupling conserves the total number of excitations. 

Next, we drive the two-level system, initially in its ground state, with 
a sequence of two resonant Gaussian pulses
\begin{align}
H_S=\frac{\hbar}2 \Omega(t) \big(|e\rangle\langle g| + |g\rangle\langle e|\big),
\end{align}
with
\begin{align}
\Omega(t)=& \frac{A_1}{2\pi \sigma_1} e^{-\frac 12(t-t_1)^2/\sigma_1^2}
          + \frac{A_2}{2\pi \sigma_2} e^{-\frac 12(t-t_2)^2/\sigma_2^2},
\end{align}
where $A_1=A_2=\pi$ are the pulse areas, $t_1=5/\kappa$ and $t_2=13/\kappa$ 
are the centers of the pulses, and
$\sigma_i=\tau_i/\sqrt{8\ln 2}$ are the standard deviations with
FWHM pulse durations $\tau_i$.
The time-dependent Rabi frequency $\Omega(t)$ is depicted in 
Fig.~\ref{fig:lorentzian}(c) 
for different values of the pulse widths $\tau_1=\tau_2=\tau$.

The corresponding two-level excitations $n_e$ and photon numbers 
$n_\textrm{ph}^\textrm{int}$ are shown in Fig.~\ref{fig:lorentzian}(d) and (e), respectively.
For short pulses with $\kappa\tau=0.1$, the two-level excitation $n_e$ 
after the first pulse closely resembles the dynamics of the initial value
calculation in Fig.~\ref{fig:lorentzian}(b).
The second pulse hits the two-level system when it is nearly in its ground
state, promoting it again to the excited state, for which the excitation is
again transferred to the photonic environment with a dynamics similar to
an initial value calculation. Thus, each pulse introduces one excitation that
is eventually converted to a photon, so the photon number 
$n_\textrm{ph}^\textrm{int}$ approaches a value of 2.

For longer pulses, we first observe an increase in the final photon number
$n_\textrm{ph}^\textrm{int}$. This is explained by the fact that excitations
are emitted as photons already within the pulse duration $\tau$, which 
facilitates the extraction of more than one excitation per
laser pulse. For even longer pulses, when the pulse width becomes comparable
to the period of excitation oscillations between system and environment, 
excitation oscillations generated at different points in time within
the pulse duration destructively interfer. 
As a result, the capabilities of the composite emitter and environment system 
to absorb excitations from the external pulse are reduced, leading to values
of $n_\textrm{ph}^\textrm{int}$ well below 2.

\subsection{Inner bonds of PT-MPOs from different algorithms: 
Example in quantum thermodynamics }

The construction of environment observable closures relies on the key property
of the ACE algorithm that the relation between inner bonds and the 
Liouville space of individual environment modes is clearly identifiable 
before MPO compression. 
Other algorithms for constructing PT-MPOs~\cite{JP,DnC,Link_infinite} 
are based on expressions for the Feynman-Vernon influence 
functional~\cite{FeynmanVernon},
where the environment degrees of freedom have been integrated out using 
path integral techniques. 
This obfuscates the relation between the inner bonds of the PT-MPOs and
the space of environment excitations. As the starting point
of these algorithms is the bath correlation function, they make no reference
to any particular discretization of the environment mode continuum.
This raises the question whether our interpretation of inner PT-MPO bonds is 
specific to ACE or applies more universally to any PT-MPO technique.

One argument in favour of the latter is the fact that the physical influence 
of the environment, and thus the Feynman-Vernon influence functional, 
should be identical for all methods that are numerically exact.
Moreover, MPO compression using SVDs provides at least locally optimal 
compression~\cite{MPS_Schollwoeck}. 
If the PT-MPO algorithms under consideration were to yield 
globally optimal compression, one would expect the resulting PT-MPOs 
to be identical, up to the intrinsic gauge freedom of MPOs.
On the other hand, comparisons have already revealed that different algorithms
may lead to PT-MPOs with different inner bond dimensions, e.g., 
due to accumulation of numerical error or truncation of MPOs in
non-normalized form~\cite{DnC}.

Here, we address this question in a numerical experiment. We calculate
PT-MPOs for an open quantum system first using ACE and then using 
the algorithm by J{\o}rgensen and Pollock (JP)~\cite{JP}, which is based on
the Feynman-Veron influence functional. 
%and is applicable to Gaussian environments like the spin-boson model. 
We calculate environment 
observable closures for the ACE PT-MPO and, after fixing the gauges,
apply these closures to the inner bonds of PT-MPOs obtained from the JP
algorithm. If the information conveyed in the inner bonds of PT-MPOs 
is universal, i.e. independent of the details of the algorithm, we
expect to find at least qualitative agreement between environment 
observables extracted from both PT-MPOs.

To remove ambiguities due to the gauge freedom, we proceed as follows:
After calculating PT-MPOs using the two algorithms, we perform additional
sweeps to ensure that the PT-MPOs are stationary. A final forward sweep
implicitly orders the basis of the inner bonds such that the rows of the 
PT-MPO matrices with respect to the inner indices are associated  
to singular values in decreasing order. Whenever for a give time step the
inner dimensions of the PT-MPOs calculated using both approaches differ,
the observable closures from ACE calculations are truncated or zero-padded
to match the bond dimensions of the JP PT-MPO. A remaining ambiguity arises from common phase factors within each row of the PT-MPO matrices, which may be 
passed on to the corresponding columns of the PT-MPO matrix of the next time
step. We fix this by extracting the phase of the largest element within
each row of an ACE PT-MPO matrix and modify the corresponding phase of the 
JP PT-MPO. 

We apply this approach to a test case in quantum thermodynamics~\cite{QThermo_Haenggi2009,QThermo_Segal2011, QThermo_IlesSmith2014,QThermo_Eisert2018,QThermo_Shubrook}.
In this research area, one investigates how concepts of macroscopic 
thermodynamics can be generalized to quantum mechanical systems. 
While work and heat play a central role in classical thermodynamics, their
definition is more complex in quantum settings~\cite{Work_observable,Rubino2022}.
Irrespective of such subtleties we are here interested in 
calculating the total energy absorbed by a quantum system subject to 
external driving. If this system is a non-Markovian open quantum system, 
a further challenge is to resolve how the total absorbed energy is distributed
over different terms in the Hamiltonian, namely the mean system energy,
the mean energy absorbed into purely environmental degrees of freedom, and 
the mean system-environment interaction energy. % which is a consequenceof non-vanishing system-environment correlations.
The latter is a consequence of non-vanishing system-environment correlations and highlights the need for methods in quantum thermodynamics that can account for strong system-environment coupling~\cite{QThermo_Haenggi2009,QThermo_Segal2011,QThermo_IlesSmith2014,QThermo_Eisert2018,QThermo_Shubrook}.
Here, the energy distribution over the different terms can be readily obtained from PT-MPOs via their inner bonds.

Concretely, we consider a two-level quantum emitter in contact with
a phonon bath described by the spin-boson model.
\begin{subequations}
\label{eq:spinboson}
\begin{align}
H=&H_S + H_E^0 + H_I +H_{PS},\\
H_E^0=& \sum_k \hbar\omega_k b^\dagger_k b_k, \\
H_I = & \sum_k \hbar g_k (b^\dagger_k +b_k) |e\rangle\langle e|,\\
H_{PS} = &  \sum_k \hbar \frac{g_k^2}{\omega_k} |e\rangle\langle e|,
\end{align}
\end{subequations}
where $H_S$ is Hamiltonian acting only on the two-level system, 
$H_E^0$ is the energy of the free phonon path, 
$H_I$ describes the system-environment interaction, and
$H_{PS}$ is added to renormalize the excited state energy such that the
polaron shift is cancelled.

We are interested in the energetics, i.e., how the energy is distributed
over time between the different terms in the total Hamiltonian.
$\langle H_S(t)\rangle$ as well as $\langle H_{PS}(t)\rangle$ 
can be directly obtained from the reduced system density matrix. 
Because system observables are directly driven by the interaction term
$H_I$, the observable closure for the environment observable
$\hat{O}_1=\sum_k \hbar g_k b^\dagger_k$ converges quickly
($\sum_k \hbar g_k b_k=\hat{O}_1^\dagger$ does not have to be calculated
separately).  
Thus, we extract the mean interaction energy $\langle H_I(t)\rangle$ via
the corresponding closure. 
Finally, the free phonon energy enters the equation of motion for the
system observables only indirectly by affecting system-bath correlations.
As discussed in the previous example, this implies that faster convergence
is expected when the increase of the free phonon energy with respect 
to its initial value 
$\langle \Delta H_E^0 (t)\rangle=
\langle H_E^{0}(t) \rangle -\langle H_E^0(0)\rangle$
is not extracted directly but instead
by integrating the equation of motion
\begin{align}
\langle \Delta H_E^0 (t)\rangle
=&%\nonumber\\ &=
-\frac{i}{\hbar} \int\limits_0^t dt' \sum_{k}\langle [H, H_E^0]\rangle_{t'} 
\nonumber\\=&
\int\limits_0^t dt' \sum_k  2\hbar \omega_k g_k \textrm{Im}\big\{
\langle \big(b^\dagger_k |e\rangle\langle e| \big) \rangle_{t'}\big\}.
\end{align}
The right-hand side is obtained by constructing the environment observable 
closure for $\hat{O}_2=\sum_k  2\hbar \omega_k g_k  b^\dagger_k$. 

First, we consider continuous resonant driving with 
$H_S=\frac{\hbar}{2} \Omega(|e\rangle\langle g|+|g\rangle\langle e|)$
with Rabi frequency $\Omega=1$ ps$^{-1}$ turned on at time $t=0$ for 
a two-level system initially in its ground state 
$\bar{\rho}=|g\rangle\langle g|$.
For the phonon environment, we assume a spectral density 
$J(\omega)=\sum_k g_k^2 \delta(\omega-\omega_k)$ of the form
$J(\omega)= \omega^3 \left(c_e e^{-\omega^2/\omega_e^2} - c_h e^{-\omega^2/\omega_h^2}
\right)^2$
with $c_e=0.1271$ ps$^{-1}$, $c_h=-0.0635$ ps$^{-1}$,
$\omega_e=2.555$ ps$^{-1}$, and $\omega_h=2.938$ ps$^{-1}$,
which is commonly used~\cite{PI_singlephoton,CoopWiercinski,PI_QRT} 
to describe longitudinal acoustic phonons interacting with
a semiconductor quantum dot in a GaAs matrix~\cite{Krummheuer}. 
We discretize $J(\omega)$ using $N=50$ modes equidistantly over the 
frequency range $[0, 7~\textrm{ps}^{-1}]$,
and assume an initial bath temperature of $T=4$ K. 
PT-MPOs are calculated using the ACE algorithm~\cite{ACE} 
for time steps $\Delta t=0.1$ ps and 
compression threshold $\epsilon=10^{-8}$.

\begin{figure}
\includegraphics[width=0.99\linewidth]{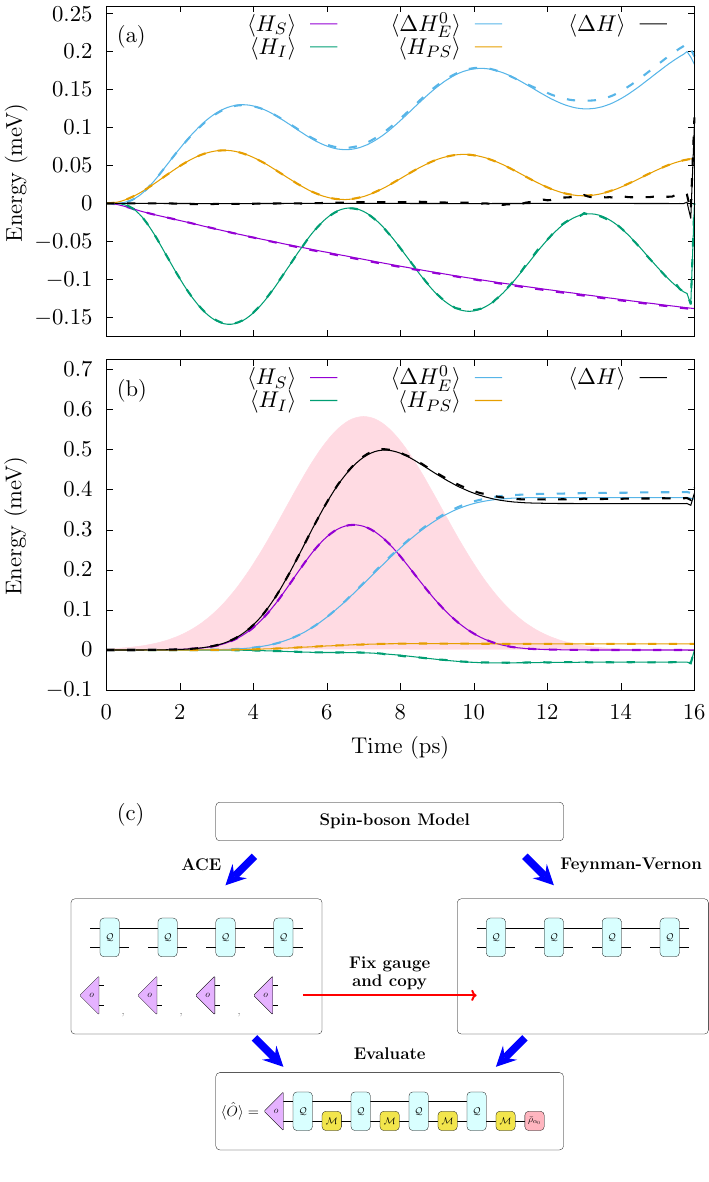}
\caption{\label{fig:spinboson}%
Time evolution of the means of the energy terms in Eq.~\eqref{eq:spinboson} for 
a quantum dot coupled to phonons under (a) resonant continuous wave driving and 
(b) excitation by a blue-detuned Gaussian laser pulse 
($|\hbar\Omega(t)|$ shaded in pink).
The symbol $\Delta$ indicates that the initial value of the free phonon energy
$\langle H_E^0(0)\rangle$ has been subtracted. Solid lines are obtained from
ACE simulations using the observable closured described in the main text for
$\epsilon=10^{-8}$ and $\Delta t=0.1$ ps. 
Dashed lines represent results of simulations, where the PT-MPOs are 
calculated using the algorithm by J{\o}rgensen and Pollock (using Feynman-Vernon path integral expressions) 
in Ref.~\cite{JP} and then transferring the observable closures obtained from the ACE simulations
as sketched in (c).
}
\end{figure}

The time evolution of the corresponding mean values of the energy terms 
in Eq.~\eqref{eq:spinboson} are shown in Fig.~\ref{fig:spinboson}(a).
The mean system energy $\langle H_S\rangle$, 
which is due to the laser driving term, starts at zero and 
slowly becomes negative as the energy is redistributed to other terms.
The polaron shift contribution $\langle H_{PS}\rangle$ is proportional to
the excited state populations and thus shows slightly damped Rabi oscillations.
The mean interaction energy is negative, which indicates binding between
the excitation and the phonon cloud, i.e., polaron formation, and 
roughly mirrors the Rabi oscillations. The free phonon energy also oscillates,
but also has an overall increasing trend, which indicates heating of the phonon bath.
The change of the total energy with respect to its initial value
$\langle \Delta H(t)\rangle = \langle H(t)\rangle - \langle H_E^{0}(0)\rangle$
remains constant because, after switching on the laser, the total Hamiltonian
is constant in time and thus energy conserving. Only in the last few time steps are
deviations found, which are again due to lack of convergence because of 
the small inner bonds at the ends of PT-MPOs.
The conservation of the total energy serves as a crucial test 
for the physicality of the results and, thus, for the suitability of our 
approach to extract all relevant energetic contributions.

With this established, we now move to a more realistic but complex process, 
the phonon-assisted excitation of a quantum dot using a blue-detuned
laser pulse. Such an excitation scheme enables bright and pure single-photon
emission with frequency separation between excitation laser and emitted
photons~\cite{PI_singlephoton}. To this end, in
Fig.~\ref{fig:spinboson}(b) we present simulations using the total 
Hamiltonian Eq.~\eqref{eq:spinboson} with system part
\begin{align}
H_S=&\frac{\hbar}{2}\big(\Omega(t) |e\rangle\langle g|
+\Omega^*(t)|g\rangle\langle e|\big)
\end{align}
and pulse envelope 
%\begin{align}
$
\Omega(t)=\frac{A}{2\pi\sigma} e^{-\frac{(t-t_0)^2}{2\sigma^2}}e^{-i\delta t},
$
%\end{align} 
and parameters $A=3\pi$, $t_0=7$ ps, $\sigma=(5\textrm{ ps})/\sqrt{8\ln 2}$, and
$\delta = 1.5\textrm{ meV}/\hbar$. Furthermore, we use the same PT-MPOs as
for Fig.~\ref{fig:spinboson}(a).

Due to the external time-dependent driving, the total energy 
$\langle H\rangle$ is no longer conserved. 
From Fig.~\ref{fig:spinboson}(b), it can be seen that
most of the absorbed energy goes into the free phonon energy,
i.e. in heating up the phonon bath. This is consistent with 
phonon-assisted excitation, where phonons are emitted during the photon
absorption process in an effectively incoherent process, while there is only
a slight build-up of system-environment correlations as measured by the
interaction energy.

Finally, the dashed lines in Fig.~\ref{fig:spinboson}(a) and (b) 
represent results, where, as
discussed above, environment observable closures from ACE calculations are 
transferred to PT-MPOs obtained by the JP algorithm. The quantities
obtained from the reduced system density matrix agree perfectly. The critical
test, the mean interaction and free phonon energies, which involve extraction
of information via inner bonds using observable closures also agree remarkably
well, even if small quantitative devations remain.
This corroborates the thesis that the information
conveyed via the inner bonds of PT-MPO, and hence the interpretation of the
inner bonds themselves, is indeed universal and not only limited to PT-MPOs
obtained from the ACE algorithm.

\section{\label{sec:discussion}Discussion}

We have shown that the inner bonds of PT-MPOs are not merely a mathematical necessity
required for describing the time-non-local memory in non-Markovian environments;
they do in fact posses a concrete physical meaning: They directly represent the 
subspace of the full environment Liouville space containing the most relevant environment
excitations, where the relevance is implicitly determined by MPO compression.
Compression via truncated singular value decomposition minimizes the compression error
without bias in favor of any particular set of system states. 
Hence, relevant environment degrees of freedom are determined without making any 
assumption about the concrete interventions on the system 
like (time-dependent) Hamiltonian evolution, Lindbladian losses, and projective measurement. 
It should be noted, however, that a more efficient representation of environmental 
influences may be available if interventions on the system are restricted.
For example, transfer tensors~\cite{TT_Cerrillo} and small matrix decomposition of path integrals~\cite{SMatPI} 
provide extremely efficient numerically exact approaches when interventions are time-translation-invariant.

Conceptionally, the connection between the inner bonds of PT-MPOs and the environment Liouville space
is expressed in terms of lossy linear transformation matrices $\mathcal{T}$ and their 
pseudoinverses $\mathcal{T}^{-1}$, both of which can be obtained by 
tracking all changes to the inner bonds at every step of the ACE algorithm~\cite{ACE}.
With the help of these matrices, one can learn information about the state of the
environment in PT-MPO simulations.
However, it turns out to be more practical to extract environment observables $\hat{O}$ 
via a set of environment observable closures $\mathfrak{o}$, 
which transform like $\mathcal{T}^{-1}$
along the ACE algorithm, and which, when applied to PT-MPO inner bonds, 
represent the effect of $\textrm{Tr}_E\{\hat{O} \rho(t)\}$.
The viability and utility of this approach is tested on a series of scenarios involving the extraction of 
environment spins, currents, radiatively emitted photons, and energies absorbed by 
time-dependently driving an open quantum system strongly coupled to an environment.

However, because the compression is lossy, the extraction of an environment observable via inner bonds
of PT-MPOs can be inaccurate if the subspace identified by MPO compression---designed to faithfully reproduce any
system observable---does not carry the information required to reconstruct the environment observable.  
In such cases, alternative methods may be useful: Observables of certain environments 
are accessible via multi-time correlation function of the system~\cite{Gribben_EnvironmentObservables}. 
More generally, environment observables can be obtained from methods 
that evolve the composite system and environment state, either by making 
physics-based approximations like in the reaction coordinate approach~\cite{Iles-Smith_PRA2014} or
by numerically exact many-body representations as in chain mapping techniques~\cite{Tamascelli_Entropy2022, Chin_env_dynamics2016,Chin_env_dynamics2017}.

On the other hand, the fact that different environment observables are extracted to different levels of
accuracy enables us to probe what information is conveyed in the inner bonds of PT-MPOs. 
In particular, we find that observables which appear earlier in the hierarchy of Heisenberg equations of motions 
starting from expectation values of system observables converge faster with respect to the MPO compression parameter.
This is in line with the fact that standard MPO compression selects environment degrees of freedom 
that most strongly affect the system evolution.

More broadly, our insights have significant consequences for fundamental and conceptual questions:
First is the pedagogical aspect: Viewing PT-MPOs as the set of environment propagators $e^{\mathcal{L}_E\Delta t}$ projected onto the (locally) most relevant environment degrees of freedom via Eq.~\eqref{eq:QTeT} is very intuitive. Yet this insight is sufficient for productively using existing PT-MPO-based open quantum systems codes~\cite{ACE,OQuPy}, and
practitioners need not understand Feynman-Vernon path integrals~\cite{FeynmanVernon}, the generalized Choi-Jamio{\l}kowski isomorphism~\cite{Choi,Jamiolkowski}, or details about matrix product states in many-body quantum physics~\cite{MPS_Schollwoeck}. 

Second, Eq.~\eqref{eq:QTeT} can facilitate a formal analysis and proofs. For example, it was
straightforward to show in Eq.~\eqref{eq:Qtwobath} that the composition of two PT-MPOs indeed
provides a numerically exact way to simulate a quantum system coupled to two environments.
A promising route for further progress is to analyze the transformations
$\mathcal{T}$ and $\mathcal{T}^{-1}$ and the subspaces to which they map. 
This could lead to novel algorithms and it could clarify connections 
to other open quantum systems methods.

Finally, it is worth stressing that Eq.~\eqref{eq:QTeT} constitutes a conceptional shift with respect to
earlier derivations of PT-MPO methods~\cite{JP}. 
The latter start from a time-non-local picture, where environment influences connecting
system states at different points in time are represented efficiently.
In contrast, in the picture developed here, the transformation matrix $\mathcal{T}^{-1}$ 
maps time-locally to the total system plus environment density matrix at a given point
in time.
Interestingly, we have also shown that environment observable closures calculated via 
ACE~\cite{ACE}, which starts from the time-local picture, can be transferred to
PT-MPOs obtained from the (time-non-local) bath correlation function~\cite{JP},
and the extracted environment observables are very similar. This suggests that
the final PT-MPOs contain the same information, i.e.~optimally compressed PT-MPOs are universal.
So, the two pictures of time-non-local memory versus time-local environment excitations
are intimately related, much like the Markov and the Born approximation tend to be
used together for modelling weakly coupled open quantum systems~\cite{BreuerPetruccione}.
While there is extensive literature on how to analyze and quantify 
non-Markovianity~\cite{nonMarkov_BLP, nonMarkov_RHP, nonMarkov_review,PT_Markov},
here, the focus on the compressed space of environment excitations suggests it
is also worthwhile to consider measures of ``non-Bornity''. The universality of PT-MPOs
further suggests that, for fixed time steps $\Delta t$, total number of time steps $n$,
and MPO compression threshold $\epsilon$, the PT-MPO bond dimension $\chi$ is a promising
candidate for such a measure of ``non-Bornity''.

\acknowledgements
This work was supported from EPSRC Grants No.~EP/T01377X/1 and No.~EP/T007214/1. 
M.C. acknowledges funding by the Return Programme of the State of North Rhine-Westphalia.
We would like to thank Jonathan Keeling, Brendon Lovett, and Gerald Fux for insightful discussions.

%\bibliography{references}
\input{inner_bonds_v2.bbl}
\end{document}

%% file: inner_bonds_v2.bbl
%apsrev4-2.bst 2019-01-14 (MD) hand-edited version of apsrev4-1.bst
%Control: key (0)
%Control: author (72) initials jnrlst
%Control: editor formatted (1) identically to author
%Control: production of article title (-1) disabled
%Control: page (0) single
%Control: year (1) truncated
%Control: production of eprint (0) enabled
%